\documentclass{iopjournal}

\usepackage{amsmath,amssymb}
\usepackage{booktabs}
\usepackage{array}
\usepackage{tikz}
\usetikzlibrary{shapes.geometric,arrows.meta,positioning,calc,fit,backgrounds}
\usepackage{multirow}
\newcommand{\etal}{\textit{et al.}}

\tikzset{
  io/.style={rectangle, rounded corners=6pt, draw=black!70, fill=blue!6,
             text width=8.4cm, align=center, inner sep=5pt, font=\normalsize},
  proc/.style={rectangle, draw=black!70, fill=gray!5,
             text width=8.4cm, align=center, inner sep=5pt, font=\normalsize},
  inner/.style={rectangle, draw=black!40, fill=orange!6,
             text width=7.6cm, align=left, inner sep=4pt, font=\small},
  loop/.style={rectangle, rounded corners=3pt, draw=black!60, dashed,
             inner sep=10pt},
  arr/.style={-{Latex[length=2.4mm]}, thick, black!70}
}

\begin{document}

\articletype{Paper}

\title{Multiscale ensemble Monte Carlo of transport and gas sensing in monolayer MoS$_2$}

\author{Lado Filipovic\orcid{0000-0002-5638-9129}}

\affil{Institute for Microelectronics, TU Wien, Vienna, Austria}

\email{filipovic@iue.tuwien.ac.at}


\begin{abstract}
Two-dimensional semiconductors, such as monolayer MoS$_2$, combine a
technologically useful band gap with a thickness of three atomic planes. The entire
carrier density then lies within a few~\AA{} of the surface, which makes them
useful as field-effect-transistor channels and as chemically sensitive devices.
Device design for MoS$_2$ needs a transport model whose scattering rates are
computed rather than calibrated, whose predictions are tested against experiment, and
that spans the scales from electron-phonon coupling to terminal current and sensor
response. We present a self-consistent ensemble Monte Carlo (EMC)
framework for monolayer MoS$_2$ built on the ViennaEMC solver, which couples a Boltzmann
transport model to Poisson's equation with degenerate free-carrier statistics. The
transport model carries wavevector-dependent acoustic, polar-optical, homopolar and
intervalley phonon scattering within and between the K valleys and the spin-orbit-split
Q valleys, together with remote substrate phonons, screened charged impurities and
surface roughness. Every intrinsic input of the transport model is computed from first principles for
this work, namely the band structure of the K and Q valleys, the phonon energies, sound
velocities and dielectric response, and the electron-phonon couplings from
density-functional perturbation theory (DFPT) matrix elements at 34 explicit phonon
wavevectors and from Wannier interpolation with the two-dimensional long-range
treatment. The model therefore contains no fitted transport parameter and no
coupling taken from another calculation, and its scattering rates are checked against
an independent solution of the linearised Boltzmann equation with the same couplings.
The EMC velocity-field characteristic feeds a one-dimensional channel model, which is
verified against a direct self-consistent three-dimensional EMC simulation of the gated
channel and reproduces the measured transfer characteristics of a CVD monolayer device
in air and vacuum. An ambient/adsorbate extension
reproduces the measured conductivity response to oxygen partial pressure and the
concentration dependence of NO$_2$ and NH$_3$ sensing of independent
monolayer devices.
\end{abstract}

\section{Introduction}\label{sec:intro}
Monolayer molybdenum disulfide (MoS$_2$) is a direct-bandgap ($\sim$1.8~eV)
two-dimensional (2D) semiconductor used as a channel material
for post-silicon field-effect transistors (FETs) and as a sensitive
platform for gas and chemical sensing~\cite{radisavljevic2011,late2013}. Its
atomic thinness gives excellent electrostatic control and exposes the entire carrier
population to the surface, but it also makes transport sensitive to phonons, the supporting dielectric, charged
impurities, roughness, and adsorbed species. A model of
device behaviour must therefore treat these mechanisms on the same physical
footing rather than absorbing them into empirical mobilities.

The ensemble Monte Carlo (EMC) method solves the semiclassical Boltzmann
transport equation (BTE) stochastically by following a statistical ensemble of
carriers through field-driven free flights and scattering events,
self-consistently coupled to Poisson's equation. Since its introduction for
hot-electron problems~\cite{kurosawa1966,fawcett1970,Kurosawa1971} and the development of the
self-scattering technique, which makes it computationally practical~\cite{rees1969},
the method has matured over five decades into a standard framework for
semiconductor transport and device simulation~\cite{jacoboni1983,price1979,jacoboni1989,selberherr1984,hess1991,moglestue1993,tomizawa1993,jungemann2003},
with full-band and technology-oriented device implementations~\cite{fischetti1988,fischetti1991}.
It is well suited to two-dimensional materials because every scattering
mechanism is described by its own microscopic rate, and hot-carrier and
non-equilibrium effects are captured without additional approximation. For monolayer MoS$_2$, the intrinsic
phonon-limited mobility and the underlying electron-phonon couplings were
established from first principles by Kaasbjerg \etal{}~\cite{kaasbjerg2012}, and the
role of the substrate and of charged impurities was clarified by Ma and
Jena~\cite{majena2014}. Experimentally, intrinsic transport was probed by
Baugher \etal{}~\cite{baugher2013} and Radisavljevi\'c and Kis~\cite{radisavljevic2013},
and the strong influence of the ambient by Ahn \etal{}~\cite{ahn2017} and Li et
al.~\cite{li2023}. In parallel, first-principles methods have matured for
computing carrier mobilities in 2D materials directly from electron-phonon
coupling~\cite{li2013,restrepo2014,gunst2016,ponce2020}, including the role
of flexural (out-of-plane) phonons enabled by the broken horizontal mirror
symmetry~\cite{fischetti2016}. These supply both the microscopic parameters and an
independent benchmark for a transport model. Here, we consider
micrometre-scale, scattering-dominated channels, in which the response is carried by the
mobility and the carrier density.

Because the entire carrier population lies at the surface, monolayer MoS$_2$ is also a
sensitive chemical transducer. This idea can be traced back to 
the detection of individual gas molecules on graphene~\cite{schedin2007}, but
the finite band gap of semiconducting transition-metal dichalcogenides removes
graphene's zero-gap limitation, yielding a much larger relative conductance
response and true field-effect operation~\cite{perkins2013,donarelli2018}.
Transduction proceeds through two coupled microscopic channels, charge
transfer between an adsorbed molecule and the channel (a donor such as NH$_3$
raises the electron density while an acceptor such as NO$_2$ depletes it), and the
additional Coulomb scattering from the resulting ionised adsorbates, which
lowers the mobility~\cite{cho2015,late2013}. Because the entire carrier density lies
within a few~\AA{} of the surface, both effects are pronounced, and room-temperature detection of NO$_2$ into
the ppb range and reversible responses to NH$_3$, humidity and volatile organics
have been demonstrated~\cite{perkins2013,late2013}. The FET geometry adds a
degree of control absent in passive chemiresistors, since a gate bias tunes the operating carrier
density and hence the selectivity and gain of the response~\cite{late2013}. The
same platform also extends to label-free biosensing~\cite{sarkar2014}. This physics
also sets the ambient baseline, as continuously adsorbed O$_2$ and H$_2$O dope and
scatter the channel, so that vacuum annealing recovers much of the intrinsic
mobility~\cite{ahn2017,li2023}. Most quantitative sensing studies remain
empirical. A transport-level description that links the microscopic
adsorbate-channel coupling to the measured device response, and that treats
sensing and intrinsic transport within the same model, is largely absent.

In this work, we combine these components into a multiscale EMC framework
for monolayer MoS$_2$. Multiscale refers to the sequence
from the first-principles inputs through the EMC transport model to the reduced device and sensor
models, each verified against the level below it. Our contributions are (i) a
complete self-consistent EMC transport model with intrinsic and extrinsic scattering mechanisms specialised to the 2D limit (Sec.~\ref{sec:emc}); (ii) a
first-principles verification of the model's electronic-structure and polar-phonon
inputs by density-functional perturbation theory (DFPT), with eleven quantities matching
the reference parametrization and experiment and the electron-phonon couplings computed
from the DFPT matrix elements (Sec.~\ref{sec:dft}, Sec.~\ref{sec:res-dft}); (iii) a hierarchical coupling
to a one-dimensional channel model, verified against a direct self-consistent EMC device
simulation, that reproduces a measured device $I_d$-$V_g$ in air and vacuum
(Sec.~\ref{sec:multiscale}, Sec.~\ref{sec:res-device}); and (iv) an ambient/adsorbate
extension reproducing measured ambient- and gas-sensing responses against independent
monolayer experiments
(Sec.~\ref{sec:sensor}, Sec.~\ref{sec:res-sensor}). Throughout, we state
which quantities are predicted and which are fitted
(Sec.~\ref{sec:discussion}).

\section{The ensemble Monte Carlo transport model}\label{sec:emc}
\subsection{Semiclassical transport} 
The framework solves the steady-state BTE for the carrier distribution
$f(\mathbf{r},\mathbf{k},t)$~\cite{lundstrom2000},
\begin{equation}
\frac{\partial f}{\partial t}
+ \mathbf{v}(\mathbf{k})\cdot\nabla_{\mathbf r} f
+ \frac{\mathbf{F}}{\hbar}\cdot\nabla_{\mathbf k} f
= \left(\frac{\partial f}{\partial t}\right)_{\!\text{coll}},
\label{eq:bte}
\end{equation}
by propagating an ensemble of $N$ superparticles. Between collisions, each carrier
undergoes a field-driven free flight governed by Newton's law,
$\hbar\,\dot{\mathbf{k}} = \mathbf{F} = -q\mathbf{E}$, so that
$\mathbf{k}(t)=\mathbf{k}(0)-q\mathbf{E}\,t/\hbar$, with the group velocity
obtained from a non-parabolic $\mathbf{k}\!\cdot\!\mathbf{p}$ valley dispersion
\begin{equation}
E(1+\alpha E)=\frac{\hbar^2 k^2}{2 m^*},
\qquad
\mathbf{v}(\mathbf{k})=\frac{1}{\hbar}\nabla_{\mathbf k}E
=\frac{\hbar\mathbf{k}}{m^*\,[1+2\alpha E]},
\label{eq:nonparab}
\end{equation}
where $m^*$ is the valley effective mass and $\alpha$ the non-parabolicity.

The stochastic free-flight duration is generated with the self-scattering
(constant-$\Gamma$) technique~\cite{rees1969,jacoboni1983,jacoboni1989}. For each valley/region, a
constant total rate $\Gamma_0$ is chosen to bound the sum of all real rates
$\lambda_i(E)$. A free flight of duration
\begin{equation}
t_f = -\frac{1}{\Gamma_0}\ln(r),\qquad r\in(0,1]\ \text{uniform},
\label{eq:freeflight}
\end{equation}
is drawn, after which a mechanism is selected by comparing a second uniform
random number against the normalised cumulative rate
$\sum_{j\le i}\lambda_j(E)/\Gamma_0$. The residual defines a null (self-)
scattering that leaves the state unchanged. Selected mechanisms update the
carrier energy and redistribute its momentum according to their differential
cross-section (isotropic, polar-angle-weighted, or intervalley as appropriate).
Rates are pre-tabulated per valley/region for efficiency, and the ensemble is
advanced over fixed time steps $\Delta t$ during which each particle may scatter
several times (Fig.~\ref{fig:flow}). The loop is parallelised over particles with
OpenMP.

The macroscopic observables follow as ensemble averages over the non-transient
part of the trajectory. The drift mobility is
\begin{equation}
\mu = \frac{\langle v_x\rangle}{E_x},
\qquad
\langle v_x\rangle=\frac{1}{N N_t}\sum_{t}\sum_{n=1}^{N} v_{x,n}(t),
\label{eq:mobility}
\end{equation}
with $N_t$ the number of averaging steps.

\subsection{Scattering mechanisms}\label{sec:mechanisms}
Table~\ref{tab:mech} lists the mechanisms included. The intrinsic mechanisms are computed for this work from the DFPT matrix elements of Sec.~\ref{sec:dft}, in the
deformation-potential forms introduced by Kaasbjerg~\etal{}~\cite{kaasbjerg2012}
extended to wavevector-dependent couplings. The extrinsic mechanisms follow Ma and Jena~\cite{majena2014} and Ridley~\cite{ridley1982}. We give the 2D forms used for the
dominant channels.

\begin{table}[ht]
\caption{Scattering mechanisms in the monolayer MoS$_2$ EMC transport model.}
\label{tab:mech}
\begin{tabular}{@{}lll@{}}
\toprule
Mechanism & Type & Origin \\
\midrule
Acoustic deformation potential, $\Xi(q)$ & intrinsic, elastic & this work\\
Zero-/first-order intervalley K-K$'$ & intrinsic, inelastic & this work\\
Intervalley K-Q, Q-Q$'$ (21 channels) & intrinsic, inelastic & this work\\
Polar-optical (2D Fr\"ohlich, LO), $|g(q)|$ & intrinsic, inelastic & this work; form of \cite{sohier2016}\\
Homopolar $A_1'$, $|g(q)|$ & intrinsic, inelastic & this work\\
Remote surface-optical phonon & extrinsic (substrate) & \cite{majena2014}\\
Screened charged impurity (2D) & extrinsic & \cite{majena2014,ridley1982}\\
Interface roughness (Prange-Nee) & extrinsic & \cite{majena2014}\\
Adsorbate Coulomb / charge transfer & extrinsic (ambient) & this work\\
\bottomrule
\end{tabular}
\end{table}

\paragraph{Valley model.}
The conduction band is represented by three valley types. The K valleys are isotropic
non-parabolic bands, Eq.~\eqref{eq:nonparab}, with $m^*=0.446\,m_0$ and
$\alpha=1.16$~eV$^{-1}$ fitted to the computed dispersion over the thermal energy range
in both in-plane directions. These values reproduce the density of states and the mean
squared velocity of the trigonally warped band to 1\,\% between 25 and 150~meV, against
a band-edge curvature mass of $0.489\,m_0$. The six Q
valleys are anisotropic non-parabolic bands ($m_l=0.646$, $m_t=1.258\,m_0$,
$\alpha=0.3$~eV$^{-1}$) whose spin-orbit splitting of 71~meV separates them, for an
electron of given spin, into two sets of three valleys $120^\circ$ apart, one at
$E_{QK}$ above the K minimum and one at $E_{QK}+71$~meV. The two sets are carried as separate valley types. A K electron scatters into the lower
set through its three nearest and into the upper set through its three farthest Q
valleys, or the reverse for the opposite spin. Spin-averaged rates represent this by
distributing each K-Q channel equally over the two sets. The separation $E_{QK}$ is
the band parameter that depends most strongly on the exchange-correlation functional and
on the geometry of the layer (Sec.~\ref{sec:res-dft}). The transport model uses
the measured 118~meV~\cite{fu2017}, and Sec.~\ref{sec:res-sens} gives the
sensitivity to it.

\paragraph{Acoustic deformation potential (elastic, equipartition).}
Acoustic phonons scatter elastically with a rate proportional to the 2D density of
states. The phonon energies exchanged by thermal electrons (3 to 5~meV at the
wavevectors $q\le2k$) are small against $k_BT$ down to 150~K, so the elastic
equipartition form applies. A monolayer deformation potential depends on the phonon
wavevector, and the transport model keeps that dependence,
\begin{equation}
\lambda_{\text{ac}}(E)=\frac{m_d\, k_B T}{\hbar^{3}\rho\, v_s^{2}}\,\bigl[1+2\alpha E\bigr]\,
\bigl\langle \Xi^2(q)\bigr\rangle_{\theta},
\qquad q=2k\sin(\theta/2),
\label{eq:acoustic}
\end{equation}
where $\Xi(q)$ is the tabulated deformation potential of the LA or TA branch, $m_d$
the density-of-states mass, $\rho$ the areal mass density, $v_s$ the sound velocity
of the branch, and the average runs over the deflection angle $\theta$, from which
the angle after scattering is also sampled. For a constant $\Xi$, this reduces to the
usual isotropic form. The Q valleys use constant values because the direct calculation
covers their couplings at two wavevectors only.

\paragraph{Polar-optical (Fr\"ohlich) and homopolar coupling.}
The longitudinal-optical (LO) branch of the $E'$ mode couples through the
two-dimensional Fr\"ohlich matrix element, which is finite at $q\to0$ and reduced at
larger $q$ by the dielectric response of the layer itself~\cite{sohier2016}. The
rate for absorption ($+$) and emission ($-$) carries the Bose occupation
$N_{\text{LO}}=[\exp(\hbar\omega_{\text{LO}}/k_BT)-1]^{-1}$,
\begin{equation}
\lambda_{\text{pop}}^{\pm}(E)=\frac{m_d A_{\text{cell}}}{2\pi\hbar^{3}}
\bigl(N_{\text{LO}}+\tfrac{1}{2}\mp\tfrac{1}{2}\bigr)
\!\int\! |g_{\text{LO}}(q)|^2\,\mathcal{S}(q,q_s)\,
\frac{q\,\mathrm{d}\theta}{k\,\sqrt{\cos^2\theta\pm\hbar\omega_{\text{LO}}/E}},
\qquad
E'=E\pm\hbar\omega_{\text{LO}},
\label{eq:frohlich}
\end{equation}
with $q=|\mathbf{k}'-\mathbf{k}|$, $A_{\text{cell}}$ the unit-cell area,
$|g_{\text{LO}}(q)|$ the tabulated first-principles matrix element of
Sec.~\ref{sec:dft}, and $\mathcal{S}(q,q_s)=[1+q_s/q]^{-2}$ the static screening by
the free carriers with the two-dimensional Thomas-Fermi wavevector $q_s$ of the sheet
density. The static form applies because the plasma energy of the sheet at the relevant
wavevectors (84 to 145~meV at $10^{12}$~cm$^{-2}$) exceeds the LO energy of 46~meV, so
the carriers follow the phonon field, and it overestimates the screening by at most 30\,\%
where the two energies come closest (Supplementary Note~S2). The homopolar $A_1'$ mode
(49~meV), whose matrix element is likewise wavevector dependent, is described by the same
expression without free-carrier screening, in the K and in the Q valleys.

\paragraph{Intervalley scattering.}
Transitions between the K and K$'$ valleys proceed through the zero-order coupling
of the $E'$ mode at K and through four first-order channels (the TA, LA, TO and
$A_1'$-like modes at K), whose matrix elements are linear in the phonon wavevector
measured from K, and the first-order intravalley TO($\Gamma$) channel is treated in the
same form. Transitions between K and Q and between Q valleys use 21 channels, each with
its own phonon energy and zero-order coupling from the DFPT calculation with the electron at
Q, assigned to the two spin sets by the selection rule of the valley model
(Supplementary Note~S2). No separate piezoelectric term is used, because in the two-dimensional
treatment the long-range part of the acoustic coupling is contained in the computed
$\Xi(q)$.

\paragraph{Extrinsic mechanisms.}
Charged impurities at the interface scatter through a two-dimensional Coulomb potential
screened by the free carriers and by the layer (Thomas-Fermi and Rytova-Keldysh), the
polar substrate contributes remote surface-optical phonon scattering with a coupling set
by its dielectric discontinuity, and interface roughness is described in the Prange-Nee form,
all in the forms of Ma and Jena~\cite{majena2014} and Ridley~\cite{ridley1982}
(Supplementary Note~S4). The adsorbate Coulomb centres of the sensor model
(Sec.~\ref{sec:sensor}) use the same screened potential.

\paragraph{Electron-phonon couplings.}
All intrinsic couplings are computed for this work (Sec.~\ref{sec:dft}), and
Supplementary Note~S2 lists every value used. The acoustic deformation potential of
the K valley is $\Xi_{\text{LA}}=6.5$~eV in the long-wavelength limit and 4.8~eV at
0.30~\AA$^{-1}$, whereas the TA coupling vanishes at $q\to0$, as it must for a
non-degenerate band, and reaches 1.4~eV at 0.30~\AA$^{-1}$, so acoustic scattering is
an LA process with a TA share of a few per cent. Obtaining this required the
two-dimensional Coulomb cutoff in the electron-phonon calculation, without which a
long-range piezoelectric-like term of the periodic slab overstates the TA coupling
several-fold (Supplementary Note~S2). The Fr\"ohlich element is 0.35~eV at $q\to0$ and
53~meV at 0.30~\AA$^{-1}$, the homopolar element 77~meV at $q\to0$, the zero-order
K-K$'$ coupling $1.7\times10^{8}$~eV/cm, and the Q valleys have $\Xi_L=3.0$ and
$\Xi_T=2.4$~eV. Compared with the LDA parametrization of
Kaasbjerg~\etal{}~\cite{kaasbjerg2012}, the LA coupling is twice as large and the
zero-order optical couplings are 0.65 to 0.8 times as large (Supplementary Note~S4). No
calibration factor is applied here. Sec.~\ref{sec:res-sens} shows that the LA coupling alone
accounts for the difference between the 392~cm$^2$/Vs obtained with the LDA couplings and
the 214~cm$^2$/Vs obtained with the computed ones. Recent first-principles values span
130 to 410~cm$^2$/Vs, depending on the treatment of the Q valleys and of the
two-dimensional Fr\"ohlich coupling~\cite{li2013,kaasbjerg2012,sohier2018,ponce2020}, and
our value lies inside that range and inside the range of measured monolayer devices.

\subsection{Self-consistent electrostatics and degenerate statistics}\label{sec:electrostatics}
\begin{figure}[ht]
\centering
\begin{tikzpicture}[node distance=4.2mm]
\node[io] (inp) {\textbf{Inputs.} Device geometry, doping $N_D/N_A$; band
   structure ($m^*$, valleys $\alpha$); scattering parameters
   (deformation potentials, Fr\"ohlich coupling, Born charges $Z^{*}$,
   $\hbar\omega_{\text{LO}}$, $v_s$), Sec.~\ref{sec:dft}};
\node[proc, below=of inp] (init) {\textbf{Initialisation.} Solve equilibrium
   nonlinear Poisson [Eq.~\eqref{eq:poisson}, Fermi-Dirac statistics of
   Supplementary Note~S4] $\to\phi_{\text{eq}}$; sample the carrier ensemble from
   $n_{\text{eq}}(\phi)$; pre-tabulate scattering rates $\lambda_i(E)$};
\node[proc, below=8.5mm of init] (pois) {Solve Poisson
   $\Rightarrow \mathbf{E}=-\nabla\phi$ \ \normalsize(frozen-field sub-cycling optional)};
\node[inner, below=of pois] (ff) {\textbf{For each particle (OpenMP):} while
   $t<\Delta t$, free-flight drift $\hbar\dot{\mathbf k}=-q\mathbf{E}$ for
   $t_f=-\Gamma_0^{-1}\ln r$ [Eq.~\eqref{eq:freeflight}]; select mechanism from
   the cumulative rate (self-scattering technique); update $(\mathbf{k},$ valley$)$};
\node[proc, below=of ff] (assign) {Assign charge to grid; inject / absorb carriers
   at ohmic \& reservoir contacts};
\node[proc, below=of assign] (stat) {If past transient, accumulate
   $\langle v_x\rangle$, current, energy statistics [Eq.~\eqref{eq:mobility}]};
\node[io, below=8.5mm of stat] (out) {\textbf{Outputs.} $\mu(n_s,T)$,
   $v_d(E)$, $\sigma$ $\to$ multiscale FET (Sec.~\ref{sec:multiscale})
   \& ambient/sensor model (Sec.~\ref{sec:sensor})};
\begin{scope}[on background layer]
\node[loop, fit=(pois)(ff)(assign)(stat), label={[font=\footnotesize\itshape]above right:time step $\Delta t$ (repeat $N_{\text{steps}}$)}] (box) {};
\end{scope}
\draw[arr] (inp)--(init);
\draw[arr] (init)--(pois);
\draw[arr] (pois)--(ff);
\draw[arr] (ff)--(assign);
\draw[arr] (assign)--(stat);
\draw[arr] (stat.east) -- ++(1.05,0) |- (pois.east);
\draw[arr] (stat)--(out);
\end{tikzpicture}
\caption{Self-consistent ensemble Monte Carlo cycle of the ViennaEMC framework~\cite{viennaemc}.
Material and scattering parameters computed by DFPT (Sec.~\ref{sec:dft}) are the inputs at the top. The dashed
block is repeated each time step, alternating a Poisson solve with a
free-flight/self-scattering update of the carrier ensemble.}
\label{fig:flow}
\end{figure}
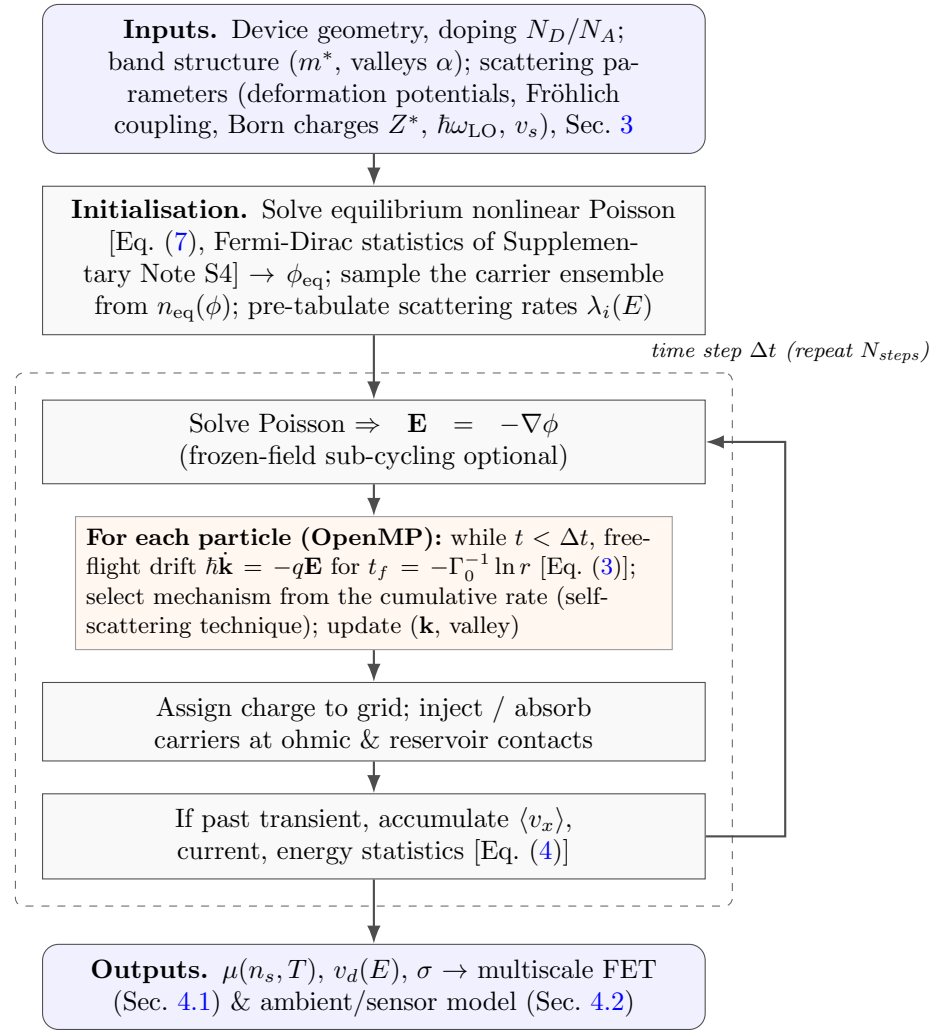

Charge is assigned to the mesh with a cloud-in-cell particle-mesh
scheme~\cite{hockney1988}, driving Poisson's equation,
\begin{equation}
\nabla\!\cdot\!\bigl(\varepsilon\,\nabla\phi\bigr)
= -q\,(p-n+N_D^{+}-N_A^{-}),
\label{eq:poisson}
\end{equation}
solved by successive over-relaxation on the device grid in the established
device-simulation manner~\cite{selberherr1984}. Because accumulation-mode 2D channels reach
degeneracy, the equilibrium density is obtained from Fermi-Dirac rather than Boltzmann statistics,
through a two-dimensional relation between the local density and the potential that
recovers the Boltzmann limit exactly for non-degenerate materials (Supplementary
Note~S4). Figure~\ref{fig:flow} summarises the complete self-consistent cycle.

\subsection{Numerical details and statistical uncertainty}\label{sec:numerics}
All transport results are ensemble averages, and every quoted value carries a
statistical uncertainty. Bulk transport (mobilities, velocity-field characteristics and
sensor transductions) is computed in a $500\times500$~nm$^2$ periodic box with a 10~nm
grid, about $2\times10^{4}$ superparticles and a time step of 0.1~fs. The temperature-dependent
mobilities average 45\,000 of 60\,000 steps after a 1.5~ps transient at a
field of 0.2~MV/m and a sheet density of $10^{12}$~cm$^{-2}$. The sensor transductions
average the second half of 25\,000 steps at each coverage, and the velocity-field
characteristics average the last three quarters of 40\,000 steps at each field. Eight
independent random seeds are used for the mobilities and the transductions, four for
the velocity-field characteristics and the sensitivity variants, and three for the
two-valley sweep of Supplementary Note~S1. The standard error is the spread over seeds
divided by the square root of their number, and it agrees with the block-averaging estimate
within one run (blocks of 0.2~ps, ten times the momentum relaxation time). Power-law
exponents are given with the standard error of the fit. The self-consistent device
simulations of Sec.~\ref{sec:multiscale} use a $5\times25\times1$~nm$^3$ grid
($53\times21\times21$ nodes), $2.4\times10^{4}$ superparticles of unit weight, a time
step of 0.1~fs and a Poisson solve every ten steps to $10^{-3}$~V. After a transient of
5~ps (6~ps at $V_{ds}=1$~V), 3~ps are averaged, and the resulting currents carry block
standard errors of 0.3 to 4\,\%.
Table~\ref{tab:stat} collects the key results with their standard errors.

\begin{table}[ht]
\caption{Key results with their statistical uncertainties. The upper part gives the
room-temperature mobility at $10^{12}$~cm$^{-2}$ (standard error over eight seeds in
parentheses) and the temperature exponent of the fit over 150 to 400~K (fit standard
error). The lower part gives the remaining quantities with the standard error over
independent seeds or, for the device currents, over blocks within one run.}
\label{tab:stat}
\small
\begin{tabular*}{\textwidth}{@{\extracolsep{\fill}}lcccccc@{}}
\toprule
 & suspended & hBN & SiO$_2$ & HfO$_2$ & Al$_2$O$_3$ & CaF$_2$\\
\midrule
$\mu$(300~K) (cm$^2$/Vs) & 212 (1) & 214 (1) & 200 (1) & 151 (2) & 176 (2) & 160 (2)\\
$\gamma$ & 1.51 (0.06) & 1.52 (0.07) & 1.64 (0.09) & 1.92 (0.10) & 1.82 (0.10) & 1.94 (0.10)\\
\bottomrule
\end{tabular*}

\medskip
\begin{tabular*}{\textwidth}{@{\extracolsep{\fill}}lll@{}}
\toprule
Quantity & Value & Standard error\\
\midrule
Low-field mobility of $v(E)$, suspended (4 seeds) & 213~cm$^2$/Vs & 4~cm$^2$/Vs\\
Drift velocity at 10~MV/m, suspended (4 seeds) & $3.55\times10^{4}$~m/s & $0.01\times10^{4}$~m/s\\
Sensitivity variants (Sec.~\ref{sec:res-sens}, 4 seeds) & & below 3\,\%\\
Direct device currents (Sec.~\ref{sec:res-device}) & & 0.3 to 4\,\%\\
Two-valley sweep (Supplementary Note~S1, 3 seeds) & & below 2\,\%\\
\bottomrule
\end{tabular*}
\end{table}

\section{First-principles calculations}\label{sec:dft}
Every intrinsic parameter of the EMC transport model is computed from first principles, namely the
band structure of the K and Q valleys, the phonon energies and sound velocities, the
Born charges and dielectric response, and the electron-phonon couplings. We use
density-functional theory and density-functional perturbation theory (DFPT) as
implemented in Quantum ESPRESSO~\cite{giannozzi2017} with the PBE
functional~\cite{perdew1996} and norm-conserving PseudoDojo
pseudopotentials~\cite{hamann2013,vansetten2018} (90~Ry cutoff and a $12\times12\times1$
mesh for the unit cell, 60~Ry and $3\times3\times1$ for the adsorbate supercells). A
projector-augmented-wave calculation~\cite{dalcorso2014} provides one of the geometries of the
K-Q sensitivity study of Sec.~\ref{sec:res-dft}.
The monolayer is treated with an in-plane-only relaxed cell and a 2D Coulomb
cutoff for the electronic and dielectric response. We unfold the in-plane dielectric response
of the slab to the cell-independent two-dimensional polarisability
$\alpha_{2D}=L(\varepsilon_\parallel-1)/4\pi$, with $L$ the cell height, which sets the
Rytova-Keldysh screening length $r_0=2\pi\alpha_{2D}$ of the impurity model of Sec.~\ref{sec:mechanisms}. Of the dielectric quantities, the transport model uses $r_0$ together with the substrate
and environment permittivities in the screened Coulomb terms, and the Born charges and
phonon energies in the polar-optical term. The slab dielectric constant itself is not
used.

The transport model includes spin-orbit coupling through the Q valleys, which split by
71~meV in our calculation (70 to 75~meV in the literature~\cite{kormanyos2015}) and are
therefore carried as two valley sets (Sec.~\ref{sec:mechanisms}). At K the conduction band splits by only 3~meV, far below $k_BT$ at the temperatures
studied, so both spin subbands are equally populated and the transport masses change by
less than 2\,\%. We therefore treat the K valley as spin degenerate. For $n$-type
transport the large valence-band splitting is irrelevant. Electron-phonon couplings are
computed in the scalar-relativistic calculation and applied to both Q sets, because
spin-orbit coupling splits
the Q valley into two spin-polarised subbands of the same orbital
composition~\cite{kormanyos2015} and therefore shifts their energies without changing
the matrix elements that couple them to the phonons. The K-Q separation itself, which
spin-orbit coupling lowers by about 34~meV, is set from the measurement
(Sec.~\ref{sec:res-dft}).

Sound velocities are the slopes of the interpolated acoustic branches between 0.04 and
0.15~\AA$^{-1}$ (Supplementary Note~S2). The electron-phonon matrix elements
$\langle\psi_{\mathbf{K}+\mathbf q}|\partial V/\partial u_{\kappa\alpha}|\psi_{\mathbf K}\rangle$
of the conduction band are computed with the Allen-Heine-Cardona
functionality~\cite{lihm2020} of the DFPT code, in a cell of height 20.2~\AA{} at the
norm-conserving relaxed lattice constant of 3.185~\AA. With the electron at K we use 21
explicit wavevectors $\mathbf q$, covering 0.03 to 0.34~\AA$^{-1}$ along $\Gamma$-M,
$\Gamma$-K and at $15^\circ$ between them, the K point for the zero-order K-K$'$
transition, six wavevectors within 0.1~\AA$^{-1}$ of K for the first-order K-K$'$
couplings, and the K-Q vector. With the electron at Q we use 13 wavevectors, covering the
intravalley directions, the three Q-Q$'$ transitions, Q-K$'$ and Q-K. This direct calculation has no two-dimensional Coulomb cutoff, so its small-wavevector acoustic couplings carry a
periodic-slab artefact. The second calculation, a Wannier interpolation with the EPW
code~\cite{giustino2017,lee2023} with the cutoff~\cite{sohier2017} and the
two-dimensional long-range treatment of Ponc\'e~\etal{}~\cite{ponce2023}, gives the
acoustic couplings along $\Gamma$-M free of it, the polar-optical and homopolar
$|g(q)|$, and the phonon-limited mobility from the linearised Boltzmann equation. We project on the phonon eigenvectors and convert to deformation potentials following
Ref.~\cite{kaasbjerg2012}. Supplementary Note~S2 gives the full tables by mode,
wavevector and direction, the comparison of the two calculations and the Boltzmann
check. Adsorbate charge transfers and adsorption energies for
Sec.~\ref{sec:sensor} come from a third set of calculations, relaxations of NO$_2$,
NH$_3$ and O$_2$ on a $4\times4$ supercell of the same layer with the D3 dispersion
correction, the two-dimensional cutoff and spin polarisation, followed by a Bader
partition of the charge density (Supplementary Note~S3). Since the reference parametrization~\cite{kaasbjerg2012} is itself a DFPT result, the
comparison of Sec.~\ref{sec:res-dft} is between two first-principles calculations.
Sec.~\ref{sec:res-sens} and Supplementary Note~S1 quantify how sensitive the transport
results are to the level of theory, in particular through the K-Q valley separation.

\section{Device and adsorbate extensions}\label{sec:extensions}

\subsection{Multiscale device model}\label{sec:multiscale}
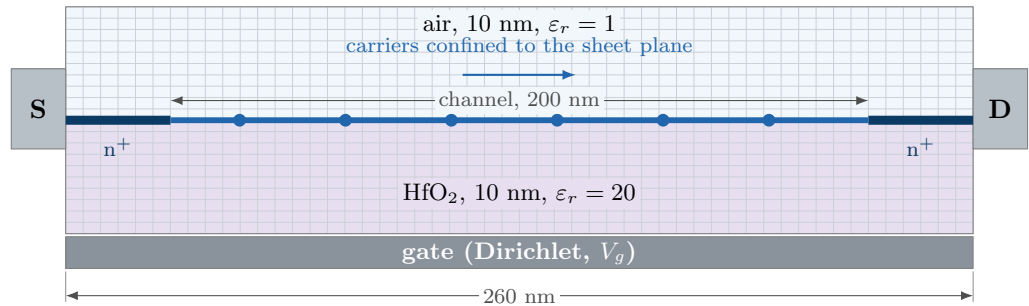
\begin{figure}[ht]
\centering
\definecolor{devair}{HTML}{F2F7FB}
\definecolor{devox}{HTML}{E7DFF0}
\definecolor{devgate}{HTML}{8A929B}
\definecolor{devcon}{HTML}{B7BFC7}
\definecolor{devmesh}{HTML}{CBD1D7}
\definecolor{devsheet}{HTML}{2166AC}
\definecolor{devnplus}{HTML}{0B3A64}
\begin{tikzpicture}[x=1cm,y=1cm,font=\small,line join=round]
\def\xa{1.385}\def\xb{10.615}
\fill[devair] (0,1.30) rectangle (12,2.80);
\fill[devox]  (0,-0.20) rectangle (12,1.30);
\foreach \i in {0,...,52} \draw[devmesh,line width=0.15pt]
  ({\i*12/52},-0.20) -- ({\i*12/52},2.80);
\foreach \j in {0,...,20} \draw[devmesh,line width=0.15pt]
  (0,{-0.20+\j*0.15}) -- (12,{-0.20+\j*0.15});
\draw[black!50] (0,-0.20) rectangle (12,2.80);
\fill[devgate] (0,-0.66) rectangle (12,-0.24);
\draw[black!50] (0,-0.66) rectangle (12,-0.24);
\node[white,font=\small\bfseries] at (6,-0.45) {gate (Dirichlet, $V_g$)};
\fill[devcon] (-0.72,0.92) rectangle (0,1.98);
\fill[devcon] (12,0.92) rectangle (12.72,1.98);
\draw[black!50] (-0.72,0.92) rectangle (0,1.98);
\draw[black!50] (12,0.92) rectangle (12.72,1.98);
\node[font=\bfseries] at (-0.36,1.45) {S};
\node[font=\bfseries] at (12.36,1.45) {D};
\draw[devnplus,line width=3.4pt] (0,1.30) -- (\xa,1.30);
\draw[devnplus,line width=3.4pt] (\xb,1.30) -- (12,1.30);
\draw[devsheet,line width=2.2pt] (\xa,1.30) -- (\xb,1.30);
\foreach \x in {2.3,3.7,5.1,6.5,7.9,9.3} \fill[devsheet] (\x,1.30) circle (2.4pt);
\node[fill=devair,inner sep=1.6pt,font=\small] at (6,2.56)
  {air, 10~nm, $\varepsilon_r=1$};
\node[fill=devox,inner sep=1.6pt,font=\small] at (6,0.32)
  {HfO$_2$, 10~nm, $\varepsilon_r=20$};
\node[font=\footnotesize,anchor=north,text=devnplus] at (0.69,1.19) {n$^+$};
\node[font=\footnotesize,anchor=north,text=devnplus] at (11.31,1.19) {n$^+$};
\node[font=\footnotesize,text=devsheet,anchor=south] at (6,2.02)
  {carriers confined to the sheet plane};
\draw[-{Latex[length=1.8mm]},devsheet,thick] (5.25,1.90) -- (6.75,1.90);
\draw[{Latex[length=1.6mm]}-{Latex[length=1.6mm]},black!70]
  (\xa,1.56) -- node[fill=devair,inner sep=1.4pt,font=\footnotesize]{channel, 200~nm} (\xb,1.56);
\draw[{Latex[length=1.6mm]}-{Latex[length=1.6mm]},black!70]
  (0,-1.02) -- node[fill=white,inner sep=1.4pt,font=\footnotesize]{260~nm} (12,-1.02);
\draw[black!45,very thin] (0,-0.70) -- (0,-1.10);
\draw[black!45,very thin] (12,-0.70) -- (12,-1.10);
\end{tikzpicture}
\caption{Structure of the direct self-consistent EMC device simulation of
Sec.~\ref{sec:multiscale}, drawn as a section along the channel with the vertical scale
exaggerated 3.25-fold. The mesh shown is the simulation mesh in this plane, 53 nodes at
5~nm along the channel and 21 nodes at 1~nm across the stack. Poisson's equation is
solved on the full three-dimensional mesh spanning
the gate oxide and the air above, whereas the carriers are confined to the single grid
plane of the monolayer, on which their charge is assigned and the in-plane force is
interpolated. The 200~nm channel is terminated by 30~nm n$^+$ regions
($3\times10^{13}$~cm$^{-2}$) that end in the ohmic source and drain reservoirs. The
channel is 500~nm wide along the direction normal to the section, meshed with a further
21 nodes at 25~nm, which completes the $5\times25\times1$~nm$^3$ cells and
$53\times21\times21$ nodes of the simulation.}
\label{fig:devgeom}
\end{figure}

The device is treated at two levels of description. The self-consistent device
simulation resolves a back-gated monolayer channel in three dimensions using ViennaEMC
(Fig.~\ref{fig:devgeom}). The gate is a
Dirichlet plane below a 10~nm HfO$_2$ oxide, the monolayer occupies one grid plane, and
10~nm of air above the sheet are included so that field lines can leave the channel. Permittivity is assigned per node (20 in the oxide, 1 in the air), and Poisson's equation
is solved with a successive over-relaxation scheme that applies the degenerate carrier
statistics of Sec.~\ref{sec:electrostatics} on the sheet nodes only. The channel is 200~nm long and
500~nm wide, contacted through 30~nm n$^+$ regions ($3\times10^{13}$~cm$^{-2}$) that end
in ohmic reservoir contacts, and the carriers are confined to the sheet plane, on which
their charge is assigned and the in-plane force is interpolated with the cloud-in-cell
scheme. Together, the n$^+$ regions and their reservoirs form an ideal ohmic termination, the
standard boundary of Monte Carlo device simulation, which isolates the channel
electrostatics from contact injection. Schottky barriers and Fermi-level pinning of
real metal contacts to MoS$_2$ lie outside the simulation, whose access resistance is
on the order of 100~$\Omega\,\mu$m, the level of the best reported semimetal
contacts, for which bismuth on monolayer MoS$_2$ gives 123~$\Omega\,\mu$m with no
Schottky barrier~\cite{shen2021}. In the comparison with measured devices they are represented by the contact
resistance of Table~\ref{tab:devpar}. The drain current is the drift-velocity (Shockley-Ramo)
tally over the channel interior,
recorded at every step after the transient, and the net contact fluxes agree with it
within 3\,\%.

Because such a simulation is too expensive for the ambient and sensing sweeps of
Sec.~\ref{sec:res-sensor}, the device response is described by a reduced model that
takes its transport input from the bulk transport model and its electrostatics from the
gradual-channel approximation, in the hierarchical sense of device
simulation~\cite{jungemann2003,selberherr1984}. The bulk EMC velocity-field characteristic
$v(E)$, computed by sweeping the driving field at the channel density
(Sec.~\ref{sec:res-device}), is used by the reduced model as a table. A common
parametrization of such a characteristic is the Caughey-Thomas form
\begin{equation}
v(E)=\frac{\mu_0 E}{\bigl[1+(\mu_0 E/v_{\text{sat}})^{\beta}\bigr]^{1/\beta}},
\label{eq:vE}
\end{equation}
with a low-field mobility $\mu_0$ and a saturation velocity $v_{\text{sat}}$. The
transport model saturates through intervalley transfer (Sec.~\ref{sec:res-device})
with a characteristic that this form reproduces only approximately, so the reduced
model integrates the tabulated $v(E)$ directly and the closed form below is used only
where the field stays in the linear range. The $v(E)$ feeds a one-dimensional gradual-channel model. The local
sheet charge follows a smooth turn-on with ideality $\eta$ (from the subthreshold
swing $SS$), giving an effective overdrive
\begin{equation}
V_{\text{ov}}(V_g)=\eta\frac{k_BT}{q}
\ln\!\Bigl[1+\exp\!\Bigl(\tfrac{q(V_g-V_{\text{th}})}{\eta k_BT}\Bigr)\Bigr],
\qquad \eta=\frac{SS}{(k_BT/q)\ln 10},
\label{eq:vov}
\end{equation}
and the current continuity $I_d=WC_{\text{ox}}\,[V_{\text{ov}}-V(x)]\,v(E(x))$ with
$\mathrm{d}V/\mathrm{d}x=E$ is integrated along the channel, the current being the value
for which the integrated length equals $L$. For the Caughey-Thomas form, this yields the
closed form with a linear-to-saturation transition,
\begin{equation}
I_d=\frac{W}{L}\,\mu_0 C_{\text{ox}}\,
\frac{V_{\text{ov}}V_d-\tfrac12 V_d^{2}}{1+V_d/(E_cL)},
\qquad
V_d=\min(V_{ds},V_{\text{dsat}}),
\label{eq:id}
\end{equation}
where $E_c=v_{\text{sat}}/\mu_0$ is the critical field and
$V_{\text{dsat}}=V_{\text{ov}}E_cL/(V_{\text{ov}}+E_cL)$, and a series contact
resistance $2R_c$ is applied self-consistently. Unlike a lumped compact model, this
resolves the channel and captures the linear-to-saturation transition of the drain
current. The reduced model has no fitted parameter when it is compared with the direct
simulation. The parameters $\mu_0$, $v_{\text{sat}}$ and $\beta$ come from the bulk EMC with the same scattering mechanisms, the channel charge $qn_s(V_g)$ from the zero-bias self-consistent
simulation, and the access resistance of the n$^+$ regions is the series term
$2R_c$. Sec.~\ref{sec:res-device} reports this verification. When a measured device is
compared, $V_{\text{th}}$, $SS$ and $R_c$ are device parameters, and so is the channel
mobility of a contact- and defect-limited CVD device, whose reduction below the intrinsic
value exceeds what charged interface impurities alone produce in the EMC
(Sec.~\ref{sec:res-sens}). Table~\ref{tab:devpar} lists every quantity of the device
model with its source.

\begin{table}[ht]
\caption{Quantities of the device model and their sources. EMC values refer to the intrinsic mechanisms at $10^{13}$~cm$^{-2}$ (Sec.~\ref{sec:res-device}), and device values refer to
the CVD monolayer FET of Ahn~\etal{}~\cite{ahn2017} in vacuum and air.}
\label{tab:devpar}
\footnotesize
\begin{tabular*}{\textwidth}{@{\extracolsep{\fill}}>{\raggedright\arraybackslash}p{4.8cm}>{\raggedright\arraybackslash}p{2.3cm}>{\raggedright\arraybackslash}p{2.0cm}>{\raggedright\arraybackslash}p{4.4cm}@{}}
\toprule
Quantity & Vacuum & Air & Source\\
\midrule
$v(E)$ & tabulated & same & EMC (Sec.~\ref{sec:res-device})\\
Low-field mobility of $v(E)$ & 213~cm$^2$/Vs & same & EMC\\
$v$ at 10~MV/m & $3.6\times10^{6}$~cm/s & same & EMC\\
$\mu_{\text{int}}$, intrinsic mobility & 213~cm$^2$/Vs & 90~cm$^2$/Vs (1~atm) & EMC, ambient model of Sec.~\ref{sec:res-sensor}\\
$\mu_0$,~channel~mobility~in~Eq.~\eqref{eq:id} & 4~cm$^2$/Vs & 1~cm$^2$/Vs & device field-effect mobility\\
$C_{\text{ox}}$ & $1.21\times10^{-4}$~F/m$^2$ & same & device geometry\\
$W$, $L$ & 32~$\mu$m, 1~$\mu$m & same & device geometry\\
$V_{\text{th}}$ & $-6$~V & 46~V & device (the shift is compared with the charge-transfer model)\\
$SS$ & 9~V/dec & 22~V/dec & device (interface traps, not modelled)\\
$R_c$ & 160~k$\Omega\,\mu$m & 500~k$\Omega\,\mu$m & device\\
$n_s(V_g)$, access resistance & & & Monte Carlo simulation (Sec.~\ref{sec:res-device})\\
\bottomrule
\end{tabular*}
\end{table}

\subsection{Ambient and adsorbate model}\label{sec:sensor}
A monolayer channel, whose carriers all lie within a few~\AA{} of the surface,
responds to adsorbates through two coupled channels:
charge transfer, which shifts the carrier density, and additional
Coulomb scattering from the ionised adsorbates, which lowers the mobility. Both
determine the EMC conductivity $\sigma=q\,n_s\,\mu$ at each fractional adsorbate
coverage $\theta$,
\begin{equation}
n_s(\theta)=n_0 \mp \Delta N\,\theta,
\qquad
\mu(\theta)=\mu\bigl[n_s(\theta),\,N_c(\theta)\bigr],
\qquad
S(\theta)=\frac{\sigma_0}{\sigma(\theta)}-1,
\label{eq:sensor}
\end{equation}
where the upper (lower) sign is an acceptor (donor), $\Delta N$ is the transferred
charge at saturation, and $N_c(\theta)=N_c^{\max}\theta$ is the areal density of
adsorbate Coulomb centres, which lower $\mu$ through the screened Coulomb scattering of Sec.~\ref{sec:mechanisms}. The
transduction $S(\theta)$ is thus an EMC output. A Langmuir isotherm
$\theta(C)=KC/(1+KC)$ maps coverage to gas concentration, with $K$ the only
per-device adsorption parameter. Oxygen partial pressure is mapped analogously.

The adsorption constant has a physical basis. Langmuir kinetics with an impingement
rate $s_0P/\sqrt{2\pi m k_BT}$ onto free sites and an Arrhenius desorption rate
$\nu N_{\text{site}}\theta\exp(-E_{\text{b}}/k_BT)$ give
\begin{equation}
K=\frac{s_0}{\nu N_{\text{site}}\sqrt{2\pi m k_BT}}\,
\exp\!\Bigl(\frac{E_{\text{b}}}{k_BT}\Bigr),
\label{eq:langmuirK}
\end{equation}
so $K$ is set by the binding energy $E_{\text{b}}=-E_{\text{ads}}$, which is positive
for a bound molecule and is taken from the adsorption energy $E_{\text{ads}}$ of
Supplementary Note~S3, together with the site density $N_{\text{site}}$, the sticking
coefficient $s_0$ and the desorption attempt frequency $\nu$. With $s_0=1$, $\nu=10^{13}$~s$^{-1}$ and $N_{\text{site}}=10^{14}$~cm$^{-2}$ the
prefactor is of order $10^{-9}$~Pa$^{-1}$. A fitted $K$ should therefore be
interpreted as an effective adsorption constant representing the unresolved distribution
of adsorption sites in a device (basal plane, vacancies, and grain boundaries), so its
device dependence reflects the defect density and the growth of each sample rather than
the transduction model (Supplementary Note~S3).

The charge-transfer amplitudes are $\Delta N=2.2\times10^{12}$~cm$^{-2}$ for NO$_2$
(acceptor) and $2.0\times10^{13}$~cm$^{-2}$ for NH$_3$ (donor) at full coverage, with
adsorbate Coulomb-centre densities $N_c^{\max}=3.0\times10^{12}$ and
$2.0\times10^{11}$~cm$^{-2}$ and a baseline density $n_0=3\times10^{12}$~cm$^{-2}$, all
fitted. To compare them with the same surface, we relaxed NO$_2$, NH$_3$ and O$_2$ on a
$4\times4$ cell of the monolayer at the geometry of the electron-phonon calculations, with the PBE
functional, the D3 dispersion correction, the two-dimensional Coulomb cutoff and spin
polarisation for the radicals, and partitioned the charge density by the Bader method
(Supplementary Note~S3, Tables~S10 and S11). All three molecules physisorb flat on the
pristine plane with adsorption energies of $-0.08$ to $-0.18$~eV and transfer 0.04 to
0.05~$e$ to NO$_2$, 0.01 to 0.03~$e$ from NH$_3$ and 0.01~$e$ to O$_2$. Per molecule, at
one molecule per $4\times4$ cell, the fitted amplitudes correspond to 0.030~$e$ for
NO$_2$, two thirds of the computed transfer, with the remainder set by the carrier limit
of the channel ($\Delta N\theta\le n_0$), and to 0.28~$e$ for NH$_3$, ten times the
pristine-plane transfer, suggesting a substantial contribution from defect-related
adsorption sites. The fitted adsorption constants support the same interpretation. Under
the assumed kinetic prefactor, Eq.~\eqref{eq:langmuirK} makes them correspond to
effective binding energies $E_{\text{b}}$ of about 0.5 to 0.6~eV, larger than the
pristine-plane physisorption energies but below the values reported for defect-modified
surfaces~\cite{kushwaha2024,kashyap2025}. This suggests an effective response involving both basal-plane and
more strongly binding defect-related sites. The published PBE transfers without dispersion
correction, 0.10~$e$ for NO$_2$ and 0.009 to 0.069~$e$ for
NH$_3$~\cite{yue2013,late2013}, bracket those given here.

\section{Results and discussion}\label{sec:results}

\subsection{Intrinsic transport and the mobility limit}\label{sec:res-transport}
\begin{figure}[ht]\centering
\includegraphics[width=0.8\textwidth]{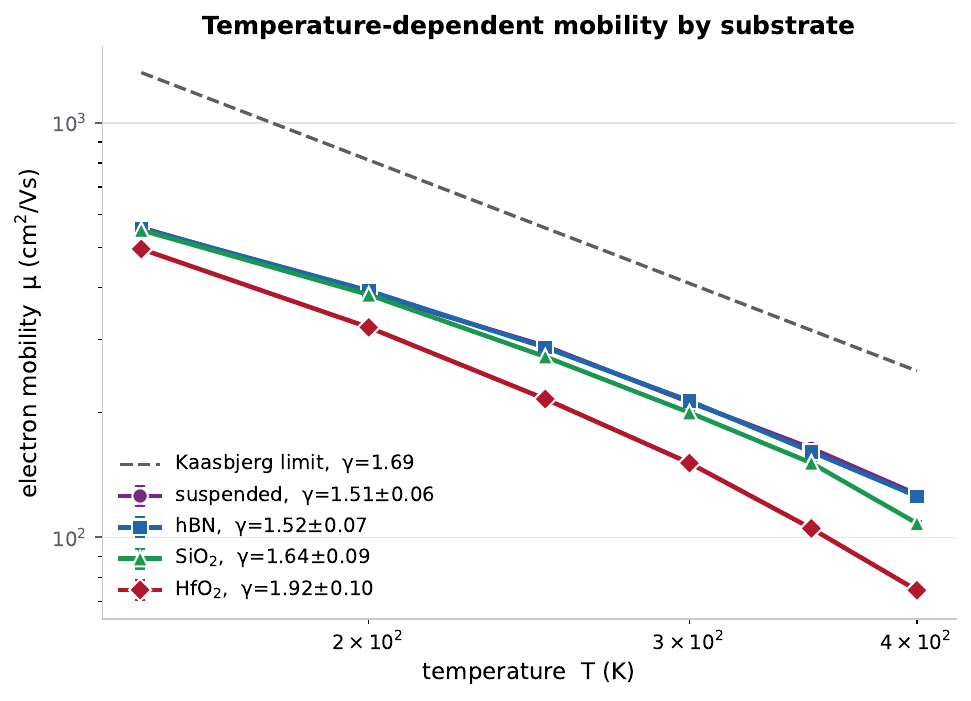}
\caption{Temperature-dependent mobility $\mu(T)$ for different substrates. Each
insulator has its own power-law exponent $\gamma$ ($\mu\propto T^{-\gamma}$, given with
the standard error of the fit). The suspended (intrinsic) film has the shallowest slope
($\gamma=1.51$). Every supporting dielectric adds a thermally activated remote
surface-optical phonon channel whose scattering grows with temperature and steepens
the decrease, most strongly for the soft, polar high-$\kappa$ HfO$_2$ ($\gamma=1.92$).
Error bars are standard errors over eight seeds and are mostly smaller than the
symbols. The impurity-free phonon limit of Kaasbjerg~\etal{}~\cite{kaasbjerg2012} ($\gamma=1.69$, dashed) is
shown for reference.}
\label{fig:transport}
\end{figure}

\begin{figure}[ht]\centering
\includegraphics[width=\textwidth]{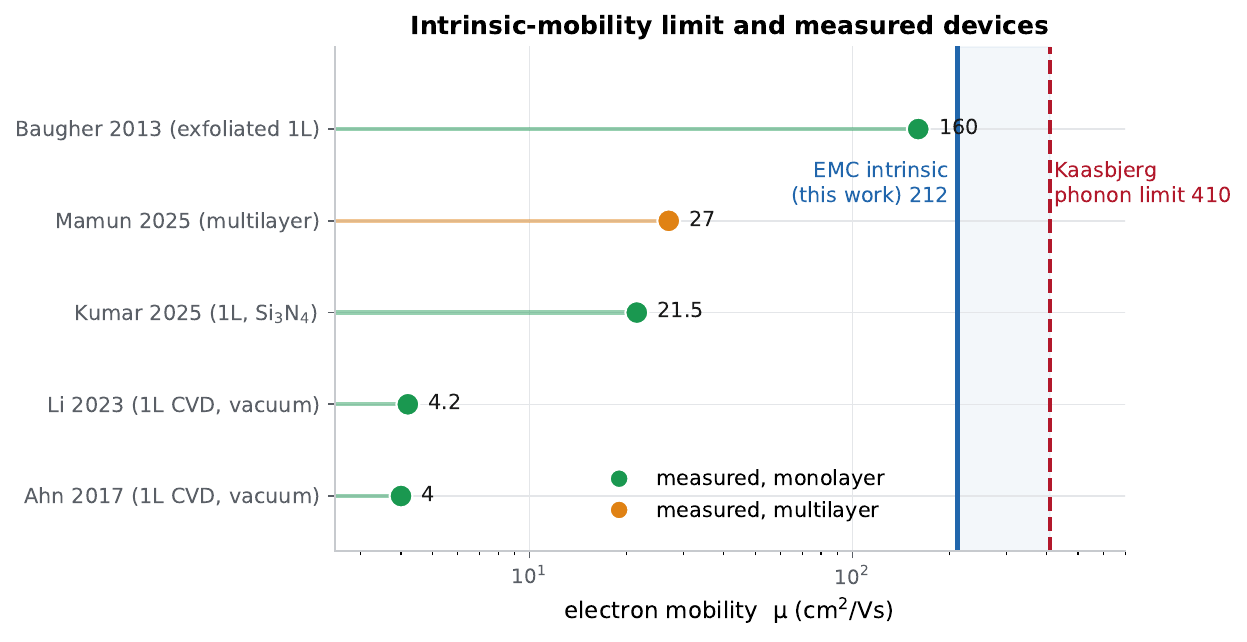}
\caption{The EMC intrinsic phonon-limited mobility sets an upper bound below the
Kaasbjerg phonon limit~\cite{kaasbjerg2012}, against which the measured monolayer and
multilayer device
mobilities~\cite{baugher2013,mamun2025,kumar2025,li2023,ahn2017} are benchmarked, each
falling below the bound by an amount that quantifies
its extrinsic (contact, defect and substrate) limitation. Monolayer and multilayer devices
are distinguished by colour, and only the monolayer devices are compared quantitatively with
the bound.}
\label{fig:bench}
\end{figure}

Figure~\ref{fig:transport} shows the computed temperature-dependent mobility for
several substrates. Each insulator follows a power law $\mu\propto T^{-\gamma}$ with
its own exponent (fit standard errors of 0.06 to 0.10). The suspended (intrinsic)
film has the shallowest slope ($\gamma\!=\!1.51$). Acoustic scattering alone gives
$\gamma=1$ in two dimensions. The optical and intervalley channels, whose Bose
factors grow with temperature, and the population of the Q valleys, which rises from
0.04\,\% at 150~K to 3\,\% at 300~K and 9\,\% at 400~K, steepen it to 1.51. Every
supporting dielectric adds a thermally activated remote surface-optical phonon channel
that steepens the decrease further, from hBN ($\gamma\!=\!1.52$) and SiO$_2$
($\gamma\!=\!1.64$) to the soft, polar high-$\kappa$ HfO$_2$ ($\gamma\!=\!1.92$). The
impurity-free phonon limit of Kaasbjerg~\etal{}~\cite{kaasbjerg2012} ($\gamma\!=\!1.69$)
lies between the suspended and the supported films. Its room-temperature value of
410~cm$^2$/Vs is twice ours because its LDA acoustic coupling is half of the computed
one (Sec.~\ref{sec:res-sens}). The room-temperature mobilities, 151 to
214~cm$^2$/Vs with standard errors of 1 to 2~cm$^2$/Vs over eight seeds, are consistent
with the best measured monolayer devices~\cite{baugher2013,radisavljevic2013}. Against an
independent Boltzmann solution with the same couplings, the transport model agrees within
1\,\% at 150~K and 13\,\% at 400~K (Supplementary Note~S2). Figure~\ref{fig:bench} recasts this as a
benchmark in which the EMC sets the intrinsic upper bound and measured devices fall
below it by an amount that quantifies their extrinsic (contact/defect/substrate)
limitation. The cleanest exfoliated device approaches this bound, while CVD and
contact-limited devices lie one to two decades lower.

\subsection{Substrate and dielectric dependence}\label{sec:res-substrate}
\begin{figure}[ht]\centering
\includegraphics[width=\textwidth]{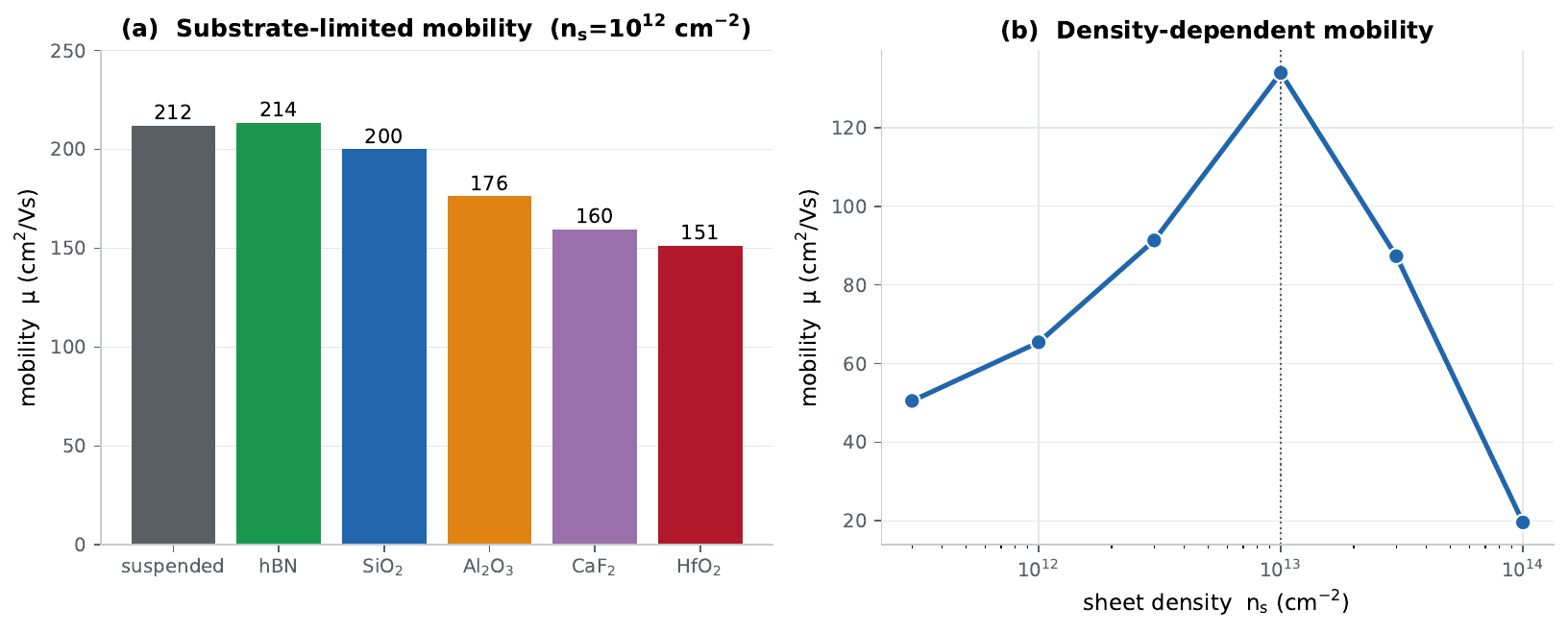}
\caption{The mobility is limited both by the supporting dielectric, shown in (a) as
the remote-surface-optical-phonon-limited mobility for the different substrates, and by
the carrier density, shown in (b) as a non-monotonic dependence $\mu(n_s)$ that is low
at small sheet density where too few free carriers screen the charged impurities, rises
as the added carriers progressively screen them, and falls again at high density as the
stronger vertical field presses the carriers into the rough dielectric interface, so
that the balance of the two opposing trends sets the peak near
$n_s=10^{13}$~cm$^{-2}$.}
\label{fig:substrate}
\end{figure}

Figure~\ref{fig:substrate}(a) shows the remote-phonon-limited mobility for each
substrate at fixed density. The ordering
suspended\,$\approx$\,hBN\,$>$\,SiO$_2$\,$>$\,Al$_2$O$_3$\,$>$\,CaF$_2$\,$>$\,HfO$_2$
is governed by a competition between the dielectric discontinuity and the phonon
stiffness. High-$\kappa$ HfO$_2$ has soft ($\sim$12~meV) modes that are
strongly thermally occupied, whereas the crystalline fluoride CaF$_2$, despite the
largest dielectric step, retains three quarters of the intrinsic mobility because its single surface-optical
mode is stiff ($\sim$51~meV) and weakly populated at room temperature, consistent
with proposed use of CaF$_2$ as a 2D-FET insulator~\cite{illarionov2019}.
Figure~\ref{fig:substrate}(b) shows the non-monotonic density dependence, rising as
free carriers screen charged impurities and falling as interface roughness takes
over, with the peak near $n_s=10^{13}$~cm$^{-2}$.

\subsection{Validation of the inputs}\label{sec:res-dft}
Table~\ref{tab:dft} lists the computed band, phonon and dielectric parameters
against the reference parametrization and experiment. Eleven quantities agree to within the expected PBE accuracy, namely the lattice
constant, the K-valley and Q-valley effective masses, the band gap, the $E'$ and $A_1'$
phonon energies, the Born effective charges, both in-plane sound velocities, the
two-dimensional polarisability and the Q-valley spin-orbit splitting. The inputs of the
transport model therefore sit at the level of theory of the reference parametrization,
while being used directly in the EMC solver. The
slab dielectric constant of 5.31 in the 20.2~\AA{} cell unfolds to
$\alpha_{2D}=6.9$~\AA{} and a screening length $r_0=43.5$~\AA, against 41 to 45~\AA{} in
the literature~\cite{berkelbach2013,laturia2018} and the 40~\AA{} used in the transport model. It
also fixes the strictly two-dimensional limit of the Fr\"ohlich coupling, 0.35~eV,
which agrees with the interpolated matrix element at the smallest wavevector within
5\,\% (Supplementary Note~S2).

The separation between the K minimum and the six Q valleys at 0.51 of $\Gamma$-K is
the input of Table~\ref{tab:dft} most sensitive to the level of theory and the
geometry. Our PBE bands give 190~meV at
the PAW-relaxed geometry and 265~meV at the norm-conserving geometry of the electron-phonon calculations, both without spin-orbit coupling. Published values range from 70~meV for the
local-density approximation at a compressed lattice constant~\cite{li2013,kormanyos2013}
to 405~meV for the HSE06 functional~\cite{kormanyos2013}, and the only direct
measurement, the pressure-driven K-$\Lambda$ crossover of Fu~\etal{}~\cite{fu2017}, gives
118~meV. Supplementary Note~S1 traces this spread to the sulphur height, the lattice
constant and spin-orbit coupling, which lowers every value by 33 to 34~meV because the Q
valley splits by 71~meV and the K valley by only 3~meV. At the experimental lattice
constant with the relaxed sulphur height and spin-orbit coupling, the separation is
113~meV, within 5~meV of the measurement. Because the couplings are insensitive to this
difference whereas the valley population is not, the transport model takes the couplings from
the DFPT geometry and the separation from the measurement, 118~meV, with its own
spin-orbit splitting of 71~meV, and Sec.~\ref{sec:res-sens} gives the sensitivity to the
separation.

All electron-phonon couplings, zero and first order, intravalley and intervalley,
polar and non-polar, are computed from the same calculations (Sec.~\ref{sec:mechanisms},
Supplementary Note~S2), so that no transport input is taken from
another calculation. The direct and interpolated couplings agree where they overlap (acoustic
couplings within 5\,\% for $q\ge0.13$~\AA$^{-1}$, homopolar coupling within 10\,\%),
and the transport model built from them reproduces the Boltzmann solution of the Wannier interpolation. The scattering rates agree within 5\,\% above 10~meV at 150, 300 and 400~K, and
the phonon-limited mobility, computed with the same valley positions and without
free-carrier screening, agrees within 1\,\% at 150~K and within 13\,\% at 400~K
(Supplementary Note~S2, Fig.~S6).

\begin{table}[ht]
\caption{Computed band, phonon and dielectric inputs of the EMC transport model (monolayer
MoS$_2$, PBE, norm-conserving calculation) compared against the reference
parametrization~\cite{kaasbjerg2012} and experiment. The electron-phonon couplings are
listed in Sec.~\ref{sec:mechanisms} and Supplementary Note~S2.}
\label{tab:dft}
\small
\begin{tabular*}{\textwidth}{@{\extracolsep{\fill}}>{\raggedright\arraybackslash}p{4.3cm}>{\raggedright\arraybackslash}p{2.8cm}>{\raggedright\arraybackslash}p{3.9cm}>{\raggedright\arraybackslash}p{2.6cm}@{}}
\toprule
Quantity & This work & Reference & Status \\
\midrule
Lattice constant $a_0$ & 3.185~\AA & 3.16~\AA & match ($+0.8$\,\%)\\
K-valley mass $m^*_K$ (band edge) & 0.489~$m_0$ (0.477 K-$\Gamma$, 0.530 K-M) & 0.48~$m_0$ & match\\
Q-valley masses $m_l/m_t$ & 0.646/1.258~$m_0$ & 0.64 to 0.66/1.21 to 1.31~$m_0$~\cite{kormanyos2015} & match\\
Band gap (direct, K) & 1.67~eV & 1.6 to 1.7~eV & match\\
$E'$ phonon (LO/TO at $\Gamma$) & 46.2~meV & $\sim$48~meV & match\\
$A_1'$ phonon & 49.1~meV & 50~meV & match\\
Born charges $Z^*_{\text{Mo}}/Z^*_{\text{S}}$ & $-0.99/{+}0.49$ & $-1.0/{+}0.5$ & match\\
Sound velocity $v_{\text{LA}}$ & 6580~m/s & $\sim$6600~m/s & match\\
Sound velocity $v_{\text{TA}}$ & 4020~m/s & $\sim$4200~m/s & match\\
2D polarisability $\alpha_{2D}$ ($r_0$) & 6.9~\AA{} (43.5~\AA) & 40 to 45~\AA~\cite{berkelbach2013,laturia2018} & match\\
Q-valley spin-orbit splitting & 71~meV & 70 to 75~meV~\cite{kormanyos2015} & match\\
Valley gap $E(Q)-E(K)$, coupling geometry & 265~meV & 118~meV (experiment~\cite{fu2017}), 70 to 405~meV (DFT spread) & geometry-sensitive\\
Valley gap, PAW geometry & 190~meV & & \\
Valley gap, $a_{\text{exp}}$ with SOC & 113~meV & & transport model uses 118~meV (Sec.~\ref{sec:res-sens})\\
\bottomrule
\end{tabular*}
\end{table}

\subsection{Sensitivity of the transport results}\label{sec:res-sens}
Table~\ref{tab:sens} quantifies how the level of theory and the couplings influence the
room-temperature mobility. Varying the K-valley transport mass over the PBE-to-$GW$
spread (0.43 to 0.53~$m_0$ around the fitted 0.446~$m_0$) changes the mobility by $+6$
and $-27$\,\% for the suspended layer and by $+6$ and $-27$\,\% on HfO$_2$. The band gap
affects the transport only through the non-parabolicity, so its PBE underestimate is immaterial for
electron transport. Raising the K-Q separation from the measured 118~meV to the 190~meV
of the PAW geometry raises the mobility by 7 and 5\,\%, and removing the spin-orbit
splitting of the Q valleys, which doubles the density of final states at the Q edge,
lowers it by 5 and 4\,\%. Replacing every computed coupling by the LDA couplings of Kaasbjerg~\etal{}~\cite{kaasbjerg2012} raises the suspended mobility to 392~cm$^2$/Vs,
close to their impurity-free value of 410~cm$^2$/Vs. Replacing only the optical couplings changes it by $-4$\,\%, and removing the first-order K-K$'$ channels by $+8$\,\%, so
the entire difference between the two first-principles values is the LA deformation
potential. Charged interface impurities on the HfO$_2$ stack reduce the mobility from
188 to 173, 148, 110, 63, 30 and 13~cm$^2$/Vs for densities of $10^{12}$,
$3\times10^{12}$, $10^{13}$, $3\times10^{13}$, $10^{14}$ and $3\times10^{14}$~cm$^{-2}$, respectively
(standard errors 0.2 to 5~cm$^2$/Vs, screening at $10^{13}$~cm$^{-2}$). Even
$3\times10^{13}$~cm$^{-2}$, the upper end of the sulphur-vacancy densities measured in
CVD films, leaves 63~cm$^2$/Vs, so the factor of about fifty between the intrinsic value
and the CVD device of Sec.~\ref{sec:res-device} is a contact and grain-boundary
limitation rather than a Coulomb-scattering effect.

\begin{table}[ht]
\caption{Sensitivity of the room-temperature mobility (cm$^2$/Vs, four seeds, standard
errors in parentheses) to the band inputs, the valley model and the electron-phonon
couplings, for the suspended layer and on HfO$_2$ at $10^{12}$~cm$^{-2}$.}
\label{tab:sens}
\small
\begin{tabular*}{\textwidth}{@{\extracolsep{\fill}}>{\raggedright\arraybackslash}p{9.6cm}ll@{}}
\toprule
Variant & Suspended & HfO$_2$\\
\midrule
Model as used (\mbox{$m^*=0.446\,m_0$}, \mbox{$E_{QK}=118$~meV}, \mbox{$\Delta_Q=71$~meV}, computed couplings) & 214 (1) & 151 (3)\\
$m^*=0.43\,m_0$ & 227 (1) & 161 (1)\\
$m^*=0.53\,m_0$ & 156 (2) & 110 (2)\\
$E_{QK}=190$~meV (PAW geometry) & 229 (5) & 158 (2)\\
$\Delta_Q=0$ (degenerate Q valleys) & 202 (3) & 145 (3)\\
Computed acoustic, LDA optical and first-order couplings & 205 (1) & 142 (1)\\
LDA couplings of Ref.~\cite{kaasbjerg2012} throughout (constant $\Xi$) & 392 (2) & 218 (3)\\
No first-order K-K$'$ channels & 232 (1) & 154 (4)\\
\bottomrule
\end{tabular*}
\end{table}

The K-Q separation affects the transport model directly through the Q-valley
population and the K-Q intervalley rates. Its consequences were mapped over the published range with the two-valley parameters of Pilotto~\etal{}~\cite{pilotto2022}
in Supplementary Note~S1 (Table~S3, Fig.~S2). From 190 to 118~meV, the low-field mobility
falls by 10\,\% and the drift velocity at 5~MV/m by 20\,\%. At 70~meV, the LDA value at
a compressed lattice constant, the mobility halves. That is the factor of two between
Li~\etal{}~\cite{li2013} and Kaasbjerg~\etal{}~\cite{kaasbjerg2012} which
Ponc\'e~\etal{}~\cite{ponce2020} trace to this parameter.
With the first-principles transport model, the same dependence appears in Table~\ref{tab:sens}.
Raising the separation to the 190~meV of the PAW geometry gains 7\,\%. Switching the
spin-orbit splitting off, which leaves a degenerate set of six valleys at 118~meV with
twice the density of final states at the Q edge, costs 5\,\% suspended and 4\,\% on
HfO$_2$.

\subsection{Device characteristics}\label{sec:res-device}
\begin{figure}[ht]\centering
\includegraphics[width=\textwidth]{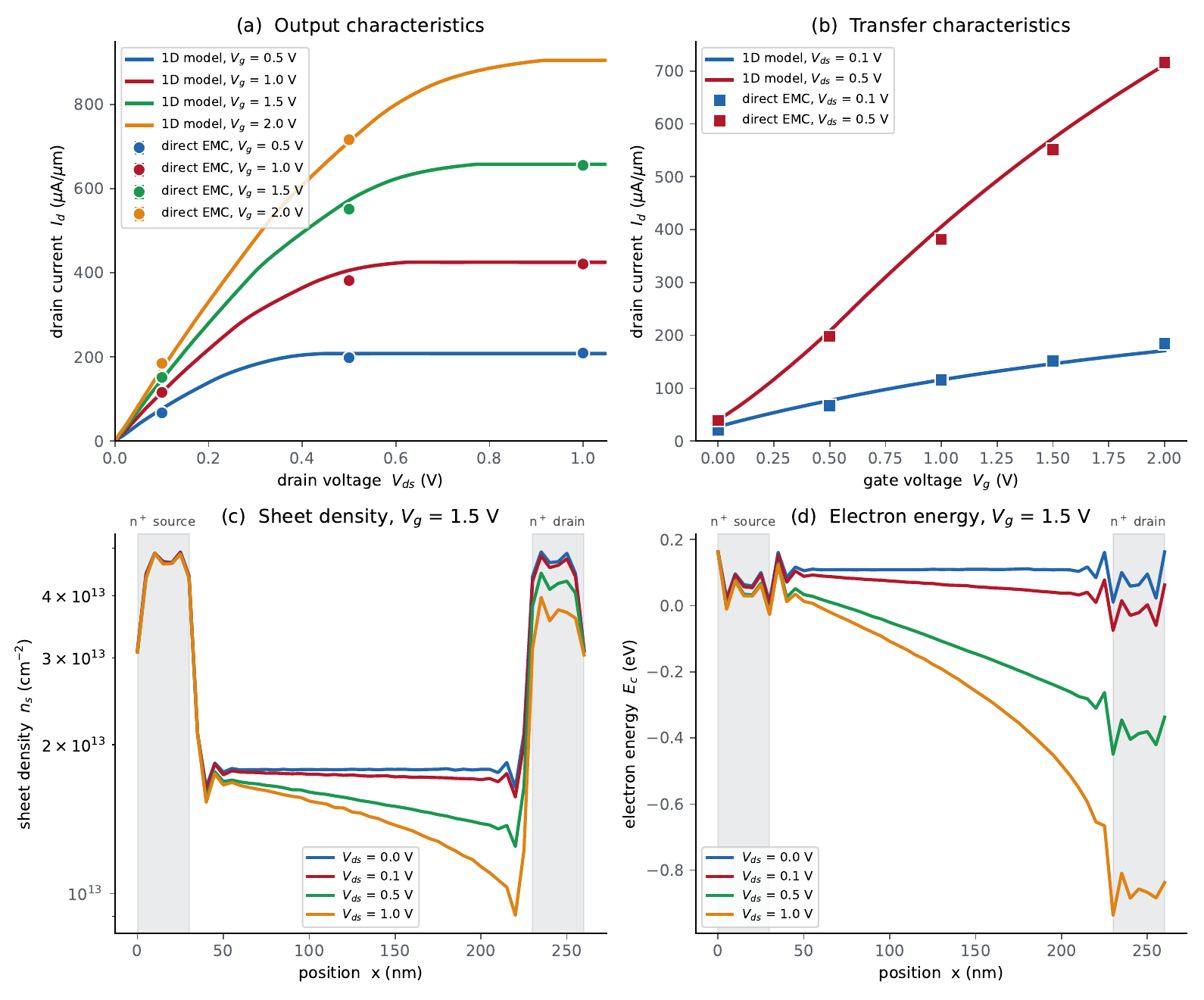}
\caption{Direct self-consistent three-dimensional EMC simulation of the gated monolayer
channel compared with the reduced one-dimensional channel model fed with the bulk EMC
$v(E)$ (no fitted parameter). (a) Output characteristics for $V_g=0.5$, 1, 1.5 and 2~V
and (b) transfer characteristics at $V_{ds}=0.1$ and 0.5~V, where symbols are the Monte Carlo simulation
with twice the block standard error, which is smaller than the symbols, and lines are the
reduced model with the tabulated EMC $v(E)$ including the access resistance of the n$^+$ regions. (c) Sheet
density and (d) electron energy along the channel at $V_g=1.5$~V for $V_{ds}=0$, 0.1,
0.5 and 1~V, where the shaded regions are the n$^+$ contact regions.}
\label{fig:direct}
\end{figure}

The reduced model is first verified against the direct ensemble Monte Carlo simulation.
Figures~\ref{fig:direct}(a) and (b) compare the two over $V_g=0$ to 2~V and
$V_{ds}=0.1$ to 1~V. The reduced model is fed with the tabulated bulk $v(E)$ of the
HfO$_2$-supported stack of the device ($\mu_0=173$~cm$^2$/Vs at the screening density of
$10^{13}$~cm$^{-2}$), the settled channel density $n_s(V_g)$ of the zero-bias runs
($1.03\times10^{13}$~cm$^{-2}$ per volt, $V_{\text{th}}=-0.26$~V) and an access
resistance of 530~$\Omega$ for the 500~nm wide device, taken as the median of the
potential drops outside the channel interior. The
two agree within 10\,\% at eleven of the thirteen bias points and within 25\,\% at all
of them. At $V_g=0$ and 0.5~V, $V_{ds}=0.1$~V, the Monte Carlo simulation lies 24 and 13\,\% below the
reduced model. There, the channel density ($2.2\times10^{12}$ and
$7.1\times10^{12}$~cm$^{-2}$) is below the screening density of the bulk $v(E)$, so the
local mobility is lower than the tabulated one (Fig.~\ref{fig:substrate}(b)). At
$V_g=2$~V, $V_{ds}=0.1$~V, it lies 8\,\% above, because the fixed access resistance
overstates the drop in the n$^+$ regions, which become more conductive with the gate. A
density-resolved table $v(E,n_s)$ and a gate-dependent access resistance would remove
most of both. Figures~\ref{fig:direct}(c) and (d) show the sheet density and electron
energy along the channel, respectively, with the gradual decrease of the density towards the drain, as
expected of a gated channel. The reduced model is therefore a verified reduced-order
form of the full simulation rather than an empirical fit, with its residual error
stated.

\begin{figure}[ht]\centering
\includegraphics[width=\textwidth]{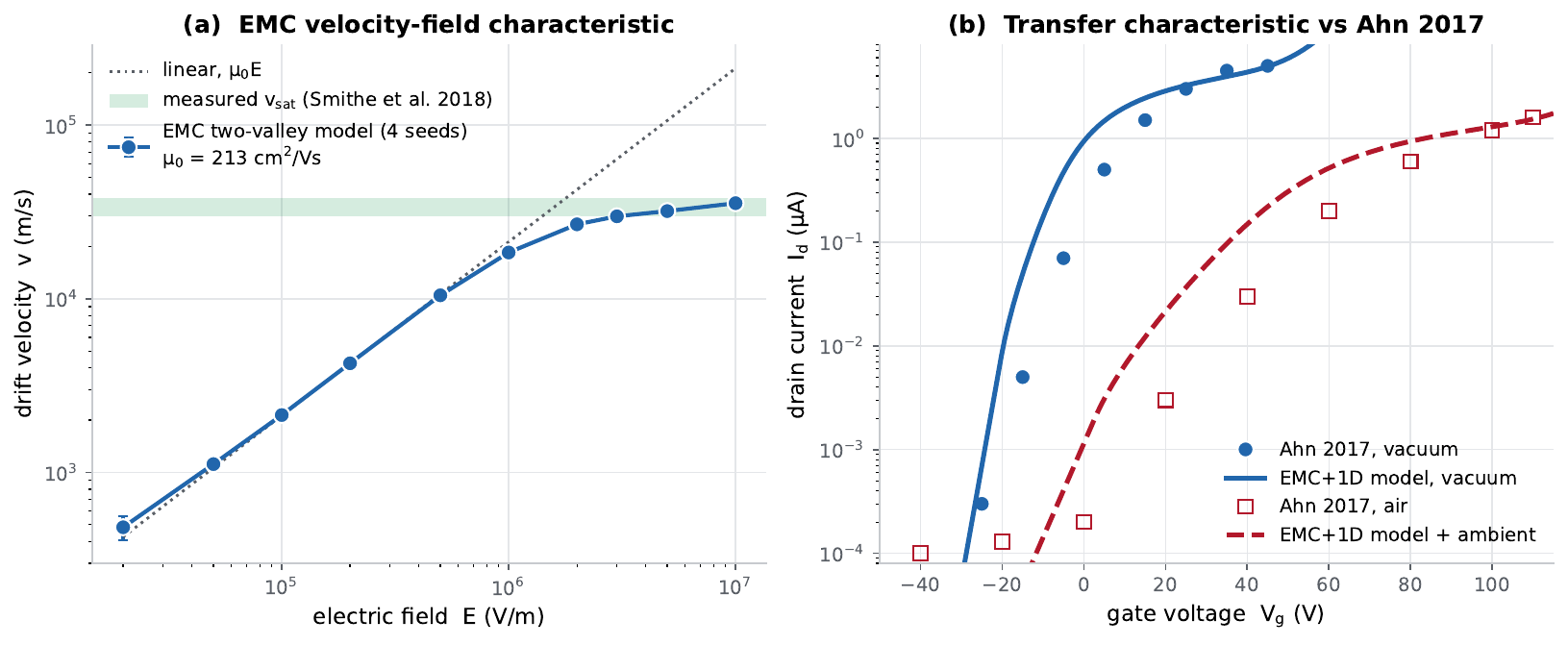}
\caption{Reduced one-dimensional channel model fed by the EMC $v(E)$.
(a) EMC velocity-field characteristic of the multivalley transport model (four seeds, standard
errors smaller than the symbols except at the lowest field), with the linear low-field relation and the measured saturation velocity of
Smithe~\etal{}~\cite{smithe2018}; the saturation above 2~MV/m is set by intervalley
transfer into the Q valleys. (b) Transfer $I_d$-$V_g$ versus the measured monolayer FET
of Ahn~\etal{}~\cite{ahn2017} in air and vacuum, with the device parameters of
Table~\ref{tab:devpar}.}
\label{fig:device}
\end{figure}

Figure~\ref{fig:device}(a) shows the EMC velocity-field characteristic $v(E)$ of the
suspended layer. It rises linearly at low field ($\mu_0=213$~cm$^2$/Vs) and bends
over above 1~MV/m into a saturation set by intervalley transfer into the heavier Q
valleys, which hold 21\,\% of the electrons at 5~MV/m and 32\,\% at 10~MV/m, where
$v=3.55\times10^{6}$~cm/s (standard error 0.01), in agreement with the measured
$(3.4\pm0.4)\times10^{6}$~cm/s~\cite{smithe2018}. The device comparisons below are made
at drain fields of at most 0.1~MV/m, where $v(E)$ is linear. Used as the input of the reduced model (Eqs.~\eqref{eq:vE} through \eqref{eq:id}), the
$v(E)$ reproduces the transfer characteristic of Fig.~\ref{fig:device}(b) against a CVD
monolayer device~\cite{ahn2017} in vacuum and air. The model reproduces both branches,
and the air$\to$vacuum change is the EMC-predicted ambient response. At 1~atm, the
ambient model of Sec.~\ref{sec:res-sensor} gives a mobility reduced by a factor of 2.4
and a carrier density reduced by 1.8, a 4.3-fold conductance change, represented in the
device model by the measured mobility ratio of 4 and a threshold shift. The absolute channel mobility of this device,
4~cm$^2$/Vs in vacuum, is two orders of magnitude below the intrinsic value
(Fig.~\ref{fig:bench}) and is a device parameter (Table~\ref{tab:devpar}). Even
$3\times10^{13}$~cm$^{-2}$ charged interface impurities leave 63~cm$^2$/Vs in the EMC
(Sec.~\ref{sec:res-sens}), so the remaining reduction is a contact and grain-boundary
limitation outside the transport model, while the velocity-saturation field and the
ambient mobility ratio are EMC predictions.

\subsection{Ambient and gas sensing}\label{sec:res-sensor}
\begin{figure}[ht]\centering
\includegraphics[width=\textwidth]{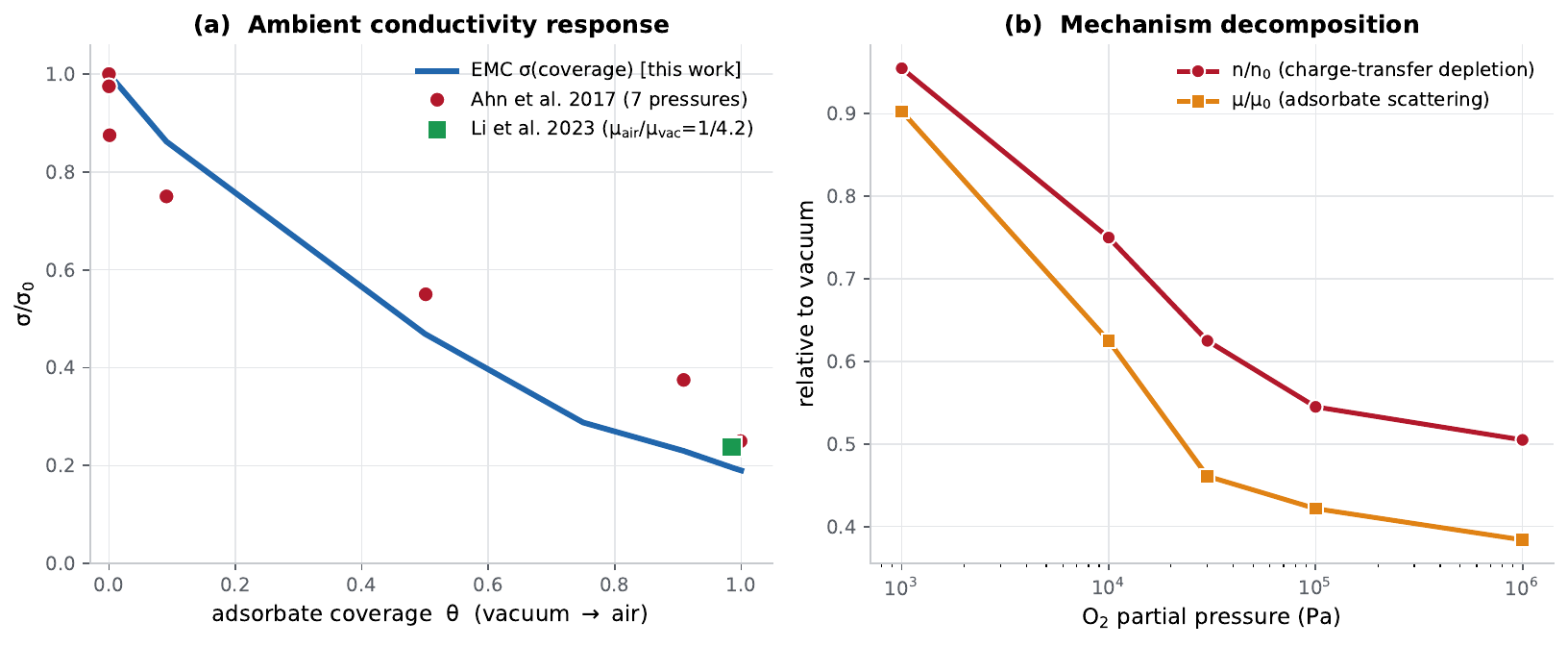}
\caption{The ambient conductivity response of the channel is shown in (a) as the EMC
$\sigma(\theta)$ versus adsorbate coverage, overlaid with measured device data from
Ahn~\etal{}~\cite{ahn2017} and Li~\etal{}~\cite{li2023}, and is decomposed in (b) into the
charge-transfer carrier-depletion and adsorbate-scattering contributions.}
\label{fig:ambient}
\end{figure}

\begin{figure}[ht]\centering
\includegraphics[width=\textwidth]{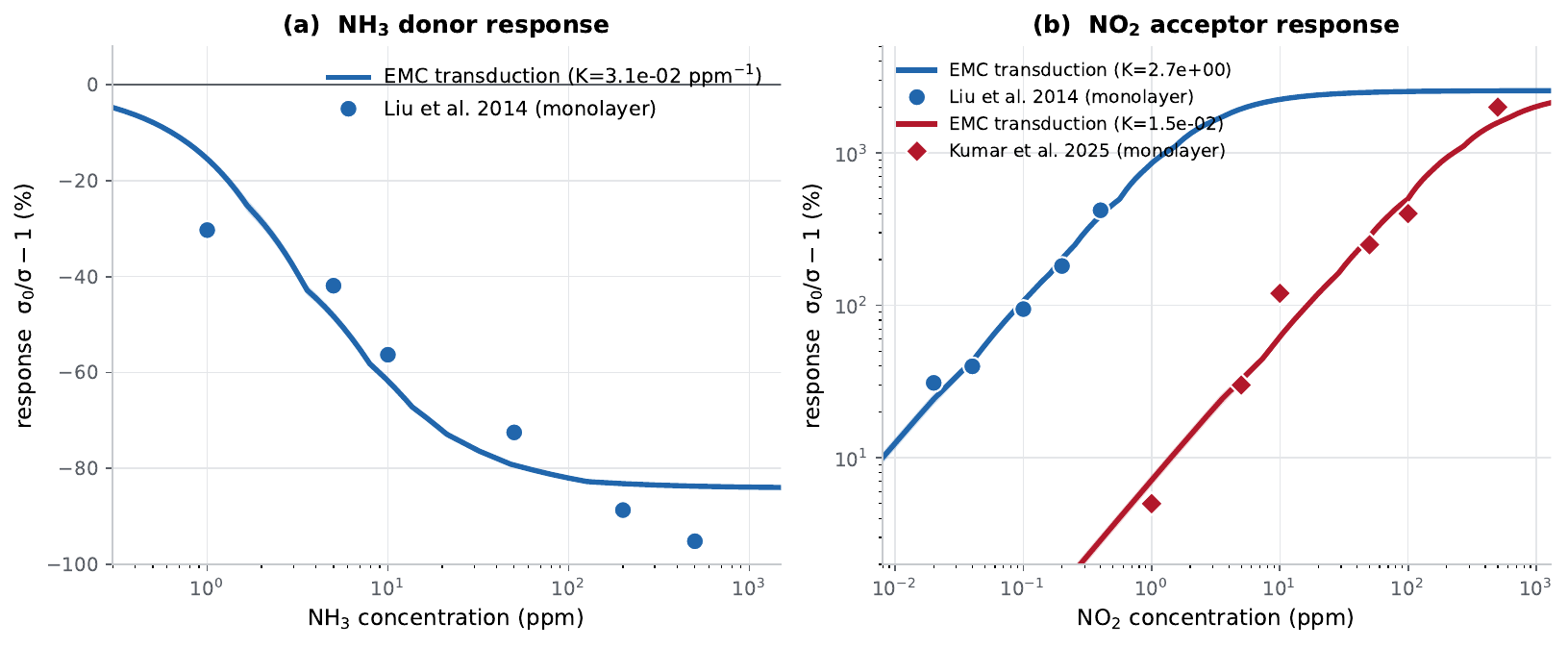}
\caption{The same EMC transduction $\sigma(\theta)$ reproduces the measured gas response
of monolayer devices, shown in (a) as the NH$_3$ donor response and in (b) as the NO$_2$
acceptor response of the CVD monolayer FET of Liu~\etal{}~\cite{liu2014}, together with
the NO$_2$ response of the monolayer device of Kumar~\etal{}~\cite{kumar2025}, each fitted
with a single per-device adsorption constant.
The standard errors of the transduction over eight seeds (2 to 9\,\% for NO$_2$, below 5\,\% for NH$_3$) are drawn as bands under the lines and are narrower than the line width.}
\label{fig:gas}
\end{figure}

\begin{figure}[ht]\centering
\includegraphics[width=.7\textwidth]{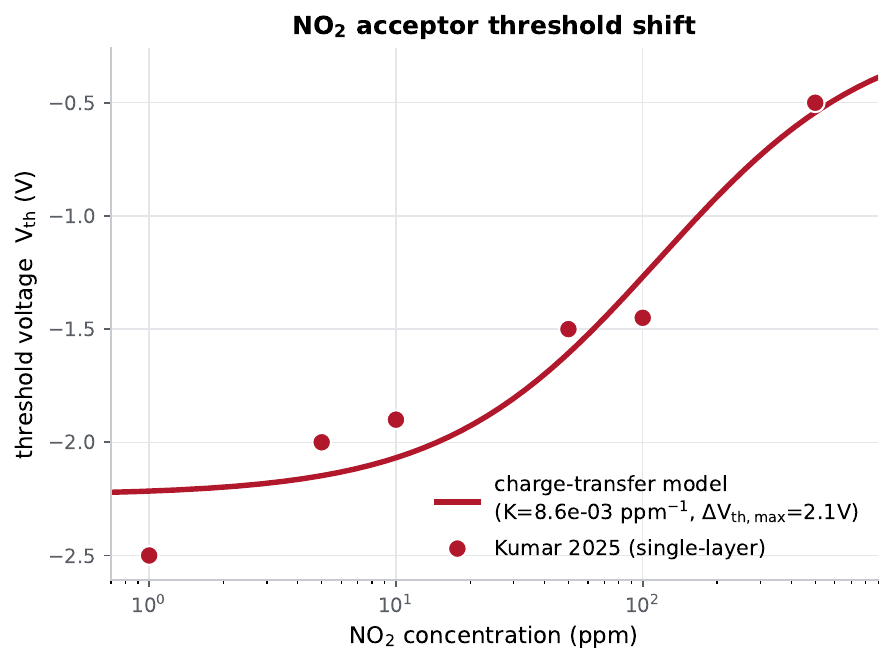}
\caption{Device-level charge transfer, the NO$_2$ acceptor threshold voltage versus
concentration of the monolayer device of Kumar~\etal{}~\cite{kumar2025}, reproduced by
the charge-transfer model as a Langmuir-saturating shift.}
\label{fig:charge}
\end{figure}

Figure~\ref{fig:ambient} shows the ambient conductivity response. The EMC
$\sigma(\theta)$ curve is populated with device measurements from Ahn~\etal{}\ (seven
pressures)~\cite{ahn2017} and Li~\etal{}~\cite{li2023}, alongside the mechanism
decomposition into charge-transfer depletion and adsorbate scattering.
Figure~\ref{fig:gas} shows the gas response against monolayer devices. The same EMC
transduction $\sigma(\theta)$ reproduces the NH$_3$ and NO$_2$ concentration dependence
of the CVD monolayer FET of Liu~\etal{}~\cite{liu2014} and the NO$_2$ response of the
monolayer device of Kumar~\etal{}~\cite{kumar2025}, with only the adsorption constant $K$
differing between devices. The NO$_2$ constants of the two devices differ by two orders
of magnitude, which Eq.~\eqref{eq:langmuirK} attributes to effective adsorption
energies of 0.47 and 0.60~eV of their site distributions.
The NH$_3$ data of Liu~\etal{}\ were taken at zero back-gate voltage, where the
baseline density is below the $3\times10^{12}$~cm$^{-2}$ of the transduction used in
Fig.~\ref{fig:gas}. At a baseline of $10^{12}$~cm$^{-2}$, the donor response is larger at
every coverage ($-59$ against $-25$\,\% at $\theta=0.05$, $-93$ against $-76$\,\% at
$\theta=0.5$, standard errors of about 1\,\%), because the same transferred charge is a
larger fraction of the channel electrons. The shape of the response is the same, so a
lower baseline rescales the fitted adsorption constant and leaves the concentration
dependence unchanged. The acceptor branch at this baseline depletes the channel above
$\theta\approx0.4$ and is not used.
Figure~\ref{fig:charge} makes the device-level comparison, where the NO$_2$
acceptor response of the same monolayer device~\cite{kumar2025} is reproduced as a
Langmuir-saturating threshold shift. Few-layer and multilayer devices, the majority of
the published sensing data, are outside the scope tested for this monolayer model.
Supplementary Note~S5 compares the monolayer transduction with a 2-layer
device~\cite{late2013} and a multilayer melamine-doped FET~\cite{mamun2025} as
qualitative comparisons, with the screening-length argument for which features of the monolayer response
apply to few-layer films and why the gate-tunable
selectivity of thick films is an interlayer effect outside the model.

\subsection{Scope and limitations}\label{sec:discussion}
Table~\ref{tab:ledger} states, per result, which quantities are EMC-predicted and
which are fitted. The transport model rests on band parameters and electron-phonon
couplings computed for this work, with no fitted transport parameter and no coupling
taken from another calculation (Sec.~\ref{sec:mechanisms}). The room-temperature
mobilities, the temperature exponents, the density peak and the substrate ordering are
predictions. The device level uses the EMC $v(E)$ and, for the verification
against the Monte Carlo simulation, no fitted parameter. For the measured CVD device the
threshold, the subthreshold swing, the contact resistance and the channel mobility are
device parameters (Table~\ref{tab:devpar}). The fitted quantities are confined to the
sensor transduction, where the per-gas charge-transfer amplitude, compared with
first-principles values in Supplementary Table~S11, and one adsorption constant per dataset,
interpreted through Eq.~\eqref{eq:langmuirK}, are fitted while the shape $S(\theta)$ is
an EMC output.

\begin{table}[ht]
\caption{Summary of predicted versus fitted quantities.}
\label{tab:ledger}
\small
\begin{tabular*}{\textwidth}{@{\extracolsep{\fill}}>{\raggedright\arraybackslash}p{3.4cm}>{\raggedright\arraybackslash}p{5.0cm}>{\raggedright\arraybackslash}p{5.4cm}@{}}
\toprule
Result & EMC-predicted (no fit) & Fitted / external \\
\midrule
$\mu(T)$ (Figs.~\ref{fig:transport},\ref{fig:bench}) & $\gamma$, absolute $\mu$ & none\\
Substrate, $\mu(n_s)$ (Fig.~\ref{fig:substrate}) & ordering, non-monotonic peak & none (substrate dielectric constants and phonon energies from the literature)\\
DFPT inputs (Table~\ref{tab:dft}, Sec.~\ref{sec:mechanisms}) & band, phonon and dielectric inputs, all electron-phonon couplings & K-Q separation from the measurement (Sec.~\ref{sec:res-dft})\\
Device I-V, Monte Carlo simulation (Fig.~\ref{fig:direct}) & $v(E)$ table, $n_s(V_g)$, access resistance, terminal currents & none\\
Device I-V, measured CVD device (Fig.~\ref{fig:device}(b)) & $v(E)$ table, saturation field, ambient conductance ratio 4.3 (measured 4) & $V_{\text{th}}$, $SS$, $R_c$ and channel mobility of the device (Table~\ref{tab:devpar})\\
Ambient (Fig.~\ref{fig:ambient}) & $\sigma(\theta)$, decomposition & pressure$\to$coverage isotherm\\
Gas (Figs.~\ref{fig:gas},\ref{fig:charge}) & transduction shape $S(\theta)$ & $\Delta N$, $N_c$ (Supplementary Table~S11), $K$ per device (Eq.~\eqref{eq:langmuirK})\\
\bottomrule
\end{tabular*}
\end{table}

The framework is developed for the monolayer, and all quantitative comparisons (the
DFPT verification in Table~\ref{tab:dft}, the transport and device results in
Figs.~\ref{fig:transport} through~\ref{fig:device}, and the sensor results in
Figs.~\ref{fig:gas} and \ref{fig:charge}) are against monolayer data. Few-layer and
multilayer devices are outside the tested scope and are compared in
Supplementary Note~S5 only qualitatively, with the physical reasons given there, and
their absolute magnitudes are not claimed as monolayer validation.

Three approximations of the transport model remain at the present level of theory. First, the
couplings are used isotropically although the direct calculation shows them to be direction
dependent at finite wavevector (the LA coupling along $\Gamma$-K is 15 to 20\,\% below
that along $\Gamma$-M), the Q-valley acoustic couplings are constant, and the K valley
is an isotropic Kane band. The comparison with the Boltzmann solution of the Wannier interpolation, which carries the full wavevector and direction dependence and the
warped band, bounds their combined effect at 5\,\% on the scattering rates and at
13\,\% on the phonon-limited mobility at 400~K (Supplementary Note~S2). Second, the K-Q
valley separation depends on the geometry and on spin-orbit coupling at the level of
tens of meV (Sec.~\ref{sec:res-dft}). The transport model uses the measured value, and
Sec.~\ref{sec:res-sens} gives the sensitivity to it. Third, the free-carrier screening
of the polar coupling is static, and the plasma-frequency argument of
Sec.~\ref{sec:mechanisms} bounds the error at 30\,\% of a channel that contributes a
fifth of the room-temperature scattering. The Boltzmann solution of Supplementary
Note~S2 carries interpolated matrix elements throughout the Brillouin
zone~\cite{giustino2017,ponce2016} and so removes the first approximation for the
phonon-limited mobility, where it bounds the difference at 8\,\% at room temperature.
Carrying that treatment through the ensemble Monte Carlo would mean abandoning the
tabulated-rate form on which the self-consistent device and sensor levels rest.

\section{Conclusion}\label{sec:conclusion}
We have presented a self-consistent, multiscale ensemble Monte Carlo
framework for monolayer MoS$_2$ that spans first-principles transport inputs, device
characteristics, and ambient and gas-sensing response. The Boltzmann transport model carries the intrinsic and extrinsic scattering
mechanisms together with a self-consistent Poisson solution and degenerate carrier
statistics. It feeds a reduced one-dimensional channel model, verified against the direct
self-consistent EMC simulation of the gated channel, and an adsorbate charge-transfer
extension. The same microscopic inputs are used from electron-phonon coupling to
terminal current and sensor transduction.

The model reproduces several independent measurements. Temperature-dependent
mobility follows a substrate-specific power law whose exponent
rises from $\gamma=1.51$ for the suspended film to $\gamma=1.92$ on HfO$_2$, as the
thermally activated remote surface-optical phonons steepen the decrease, and the
mobilities lie below the Kaasbjerg phonon limit~\cite{kaasbjerg2012}. The room-temperature mobility of
151 to 214~cm$^2$/Vs and the non-monotonic density dependence agree with the best measured
monolayer devices, and the intrinsic value provides an upper bound for
the extrinsic limitation of each measured device. The band, phonon and dielectric
inputs, computed by DFPT, agree with the reference parametrization and experiment in
eleven independent quantities, and the sensitivity of the results to the level of theory, in particular
through the K-Q valley separation, is reported. Without a fitted parameter, the reduced device model reproduces the direct EMC
simulation within 10\,\% at eleven of thirteen bias points, and with the device
parameters of the measurement it reproduces the transfer characteristic of a CVD
monolayer device in air and in vacuum. The adsorbate extension reproduces the ambient
conductivity, the oxygen-pressure response and the NO$_2$ and NH$_3$ sensing of
monolayer devices. The separation of
predicted and fitted quantities shows that the transport model uses couplings
computed from the DFPT matrix elements without a fitted transport parameter, with the
remaining fitting confined to the sensor transduction, whose amplitudes are compared
with the charge transfers and adsorption energies computed for the same surface.

The framework can be used for modelling 2D-material
field-effect transistors and chemical sensors. Every intrinsic input of the transport
model comes from first-principles calculations of the same material at one level of
theory, is checked against an independent Boltzmann solution with the same couplings,
and is carried without further adjustment to the device and sensor level.

\funding{
This research was funded in whole, or in part, by the Austrian Science Fund (FWF)  \href{https://doi.org/10.55776/P35318}{10.55776/P35318} and \href{https://doi.org/10.55776/DOC142}{10.55776/DOC142}.}

\roles{L.F. conceived the study, developed the ViennaEMC extensions and simulation
drivers, performed the transport, device, sensor and first-principles calculations,
analysed the results, and wrote the manuscript.}

\data{The simulation data supporting this study (ensemble Monte Carlo outputs with
their seeds, the bulk and device input decks, the first-principles DFPT inputs, and the
band-structure and two-valley sweep data of Supplementary Note~S1, the
electron-phonon matrix elements and Boltzmann-solution files of Supplementary Note~S2
and the adsorbate calculations of Supplementary Note~S3) are openly available
in the TU Wien Research Data repository at
\href{https://doi.org/10.48436/87zq1-28246}{10.48436/87zq1-28246}~\cite{mos2dataset}. The
ViennaEMC framework~\cite{viennaemc} is available in the project repository.}

\section*{Supplementary data}
\noindent Supplementary Notes~S1 (the K-Q valley separation, its published values, geometry
and spin-orbit sensitivities, and the two-valley transport bound), S2 (the
first-principles electron-phonon couplings of the K and Q valleys and the comparison of
the transport model with the Boltzmann solution of the Wannier interpolation), S3 (the adsorbate
calculations and the sensor-model amplitudes), S4 (the extrinsic scattering forms, the
degenerate carrier statistics and the comparison with the reference parametrization) and S5 (the
comparison with few-layer devices) follow the reference list.

\section*{Ethics statement}
\noindent This work is a computational and theoretical study based on simulation and
first-principles calculation. It involved no research on human participants, no human
data or tissue, and no animals, and therefore required no ethics approval.

\section*{Conflict of interest}
\noindent The author declares no competing interests.

\bibliographystyle{iopart-num-doi}
\bibliography{refs}

\clearpage
\setcounter{section}{0}
\setcounter{subsection}{0}
\setcounter{figure}{0}
\setcounter{table}{0}
\setcounter{equation}{0}
\renewcommand{\thesection}{Supplementary Note S\arabic{section}}
\renewcommand{\thesubsection}{S\arabic{section}.\arabic{subsection}}
\renewcommand{\thefigure}{S\arabic{figure}}
\renewcommand{\thetable}{S\arabic{table}}
\renewcommand{\theequation}{S\arabic{equation}}
\setlength{\tabcolsep}{2pt}
\begin{center}
{\large\bfseries Supplementary material}\\[2pt]
{\normalsize Multiscale ensemble Monte Carlo of transport and gas sensing in monolayer MoS$_2$}
\end{center}
\vspace{4mm}

\noindent Note~S1 concerns the conduction-band
K-Q valley separation of monolayer MoS$_2$, the input of the transport model that
depends most strongly on the exchange-correlation functional and on the geometry of the
layer. It collects the published
values, traces their spread to the lattice constant, the sulphur height and spin-orbit
coupling with PBE calculations at fixed and relaxed geometries, and quantifies the
transport consequences of the remaining uncertainty with a two-valley ensemble Monte Carlo transport model. Note~S2 documents the first-principles electron-phonon couplings of the K
and Q valleys. It gives the matrix elements of the direct DFPT calculation by phonon mode,
wavevector and direction, identifies the periodic-slab artefact in its small-wavevector
acoustic couplings, gives the zero- and first-order intervalley couplings and the polar
and homopolar couplings from the Wannier interpolation with the two-dimensional
long-range treatment, lists the values used in the transport model, and compares the scattering
rates and the phonon-limited mobility of the transport model with the Boltzmann solution of the Wannier interpolation. Note~S3 reports the adsorption energies and charge transfers of
NO$_2$, NH$_3$ and O$_2$ on the monolayer and compares them with the amplitudes of the
sensor model. Note~S4 gives the forms of the extrinsic scattering mechanisms, the
degenerate carrier statistics of the Poisson equation and the comparison of the computed couplings
with the reference parametrization. Note~S5 compares the monolayer transduction with
few-layer devices qualitatively.

\section{The K-Q valley separation}\label{sec:s1}

The conduction band of monolayer MoS$_2$ has its minimum at the K and K$'$ points and a
second set of six minima, the Q (also called $\Lambda$ or $\Sigma$) valleys, at about half
of the $\Gamma$-K distance. Their energy above the K minimum, $E_{QK}=E(Q)-E(K)$, sets the
Q-valley population at a given temperature and the onset of K$\to$Q intervalley scattering.
Of all inputs of the transport model (Table~4 of the main text), this separation is the
one that depends most strongly on the exchange-correlation functional and on the
geometry of the layer. The main text reports the conclusions of this analysis, and this
note gives the full account.

\subsection{Published values}\label{sec:s1-lit}

Table~\ref{tab:s1} collects the published values of $E_{QK}$, which span 70 to 405~meV. The only published 
direct measurement that we are aware of is the pressure-driven K-Q crossover of Fu~\etal{}~\cite{fu2017}, who
followed the direct K-K and the indirect Q-K photoluminescence lines of a monolayer
under hydrostatic pressure, found them to shift at $+49.4$ and $-15.3$~meV/GPa and to cross
at 1.9~GPa, and so obtained $E_{QK}=118$~meV at ambient pressure. Angle-resolved
photoemission on alkali-intercalated monolayers gives a lower bound of about
60~meV~\cite{eknapakul2014,kormanyos2015}.

Three trends can be extracted from the table. First, the separation decreases with
decreasing lattice constant. The local-density approximation, which contracts the lattice
to 3.13~\AA, gives the smallest separations, and Korm\'anyos~\etal{}~\cite{kormanyos2015}
attribute the difference between their two calculations to the lattice constants used.
Second, the separation depends on the functional. The HSE06 hybrid functional raises the
Q valley by several hundred meV~\cite{kormanyos2013}, whereas converged GW corrections
change the separation much less than they change the band gap~\cite{kormanyos2015}.
Third, spin-orbit coupling lowers the separation by about half of the Q-valley splitting,
because the conduction band splits by 70 to 75~meV at Q and by only 3~meV at
K~\cite{kormanyos2015}. Sohier~\etal{}~\cite{sohier2018} add the pseudopotential and the
doping level~\cite{brumme2015} to this list. The same spread appears in the calculated
mobilities. Li~\etal{}~\cite{li2013} obtain a phonon-limited room-temperature mobility of
130~cm$^2$/Vs with $E_{QK}=70$~meV and 320~cm$^2$/Vs when the Q valleys are excluded,
Kaasbjerg~\etal{}~\cite{kaasbjerg2012} obtain 410~cm$^2$/Vs in a K-only calculation with
$E_{QK}\approx200$~meV, and Ponc\'e~\etal{}~\cite{ponce2020} identify this parameter as
the main source of the discrepancy between first-principles mobilities of MoS$_2$.

\begin{table}[ht]
\caption{Published values of the K-Q separation of monolayer MoS$_2$. $a$ is the lattice
constant of the calculation, SOC denotes spin-orbit coupling.}
\label{tab:s1}
\begin{tabular}{@{}lllll@{}}
\toprule
Method & $a$ (\AA) & SOC & $E_{QK}$ (meV) & Reference \\
\midrule
LDA, LDA geometry & 3.129 & no & 90 & Korm\'anyos~\etal{}~\cite{kormanyos2013}\\
LDA, ultrasoft & 3.13 & no & 70 & Li~\etal{}~\cite{li2013}\\
LDA, PAW & 3.14 & no & $\sim$200 & Kaasbjerg~\etal{}~\cite{kaasbjerg2012}\\
LDA bands, HSE geometry & 3.157 & yes & 207 & Korm\'anyos~\etal{}~\cite{kormanyos2015}\\
PBE, PBE geometry & $\sim$3.18 & yes & 246 & Korm\'anyos~\etal{}~\cite{kormanyos2015}\\
HSE06 & HSE & no & 405 & Korm\'anyos~\etal{}~\cite{kormanyos2013}\\
PBE, PAW geometry & 3.181 & no & 190 & this work (Table~\ref{tab:s2})\\
PBE, geometry of the electron-phonon calculations & 3.185 & no & 265 & this work (Note~S2)\\
PBE, $a_{\text{exp}}$, relaxed, SOC & 3.160 & yes & 113 & this work (Table~\ref{tab:s2})\\
Experiment, pressure crossover & & & 118 & Fu~\etal{}~\cite{fu2017}\\
Experiment, ARPES lower bound & & & $\gtrsim60$ & Eknapakul~\etal{}~\cite{eknapakul2014}\\
\bottomrule
\end{tabular}
\end{table}

\subsection{Sensitivity to geometry and spin-orbit coupling}\label{sec:s1-test}

\begin{figure}[ht]\centering
\includegraphics[width=0.75\textwidth]{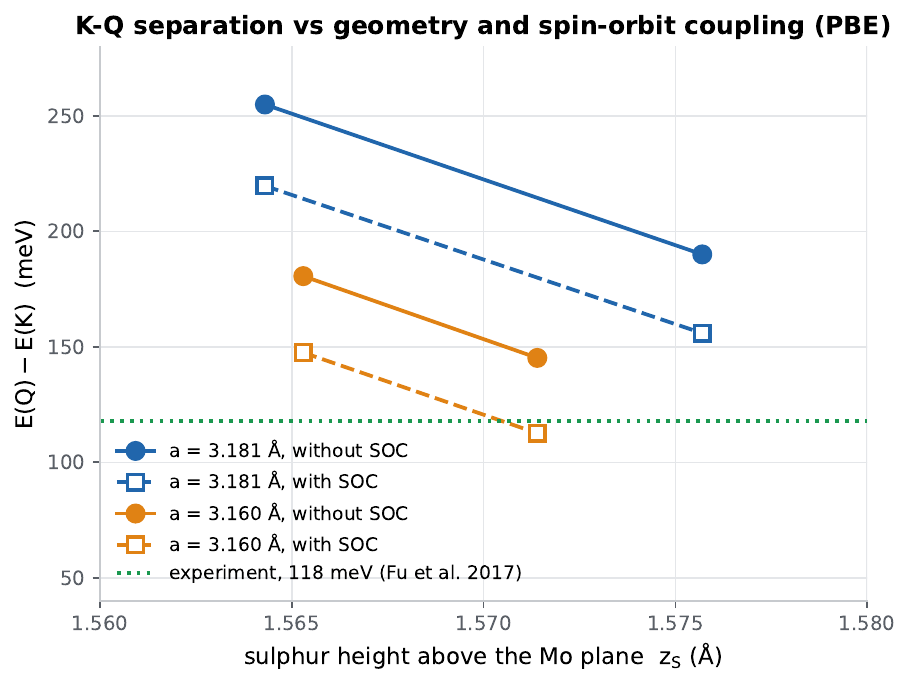}
\caption{K-Q separation from PBE as a function of the sulphur height at the two lattice
constants, without (filled circles, solid lines) and with (open squares, dashed lines)
spin-orbit coupling, data of Table~\ref{tab:s2}. The dotted line is the measured value of
Fu~\etal{}~\cite{fu2017}.}
\label{fig:s1}
\end{figure}

To separate these trends, we recomputed the separation with one functional (PBE) at fixed
and relaxed geometries, with and without spin-orbit coupling. The calculations use
Quantum ESPRESSO~\cite{giannozzi2017} with the norm-conserving scalar-relativistic and
fully relativistic PseudoDojo pseudopotentials~\cite{hamann2013,vansetten2018}, a
plane-wave cutoff of 60~Ry (240~Ry for the density), a $12\times12\times1$ k-point mesh, the
two-dimensional Coulomb cutoff and a cell height of $6.33\,a$. For the sulphur height $z_S$ above the Mo plane we either adopted the projector-augmented-wave (PAW)
relaxation used for the parameter validation of Sec.~3 of the main text
($z_S=1.576$~\AA{} at $a=3.181$~\AA) or
relaxed with the norm-conserving pseudopotentials at fixed lattice constant (forces below
$10^{-4}$~Ry/bohr). The bands were computed on an explicit list of wavevectors that
covers the K minimum and its neighbourhood, from which the effective masses follow, and
the segment between 0.40 and 0.60 of K-$\Gamma$ in steps of 0.01, on which the Q minimum
was located and refined by a parabola through the three lowest points.
Table~\ref{tab:s2} lists the eight cases and Fig.~\ref{fig:s1} plots them.

\begin{table}[ht]
\caption{K-Q separation from PBE at fixed and relaxed geometries, with and without
spin-orbit coupling. $z_S$ is the sulphur height above the Mo plane, and PAW and NC (norm-conserving) mark the
pseudopotential family with which it was relaxed, ``PAW fr.'' the PAW fractional
coordinate transferred rigidly to the smaller lattice constant. $q_Q$ is the position of the
Q minimum along $\Gamma$-K in units of the $\Gamma$-K distance, $m^*_K$ the K-valley mass
along K-$\Gamma$ and K-M, $m^*_{Q,l}$ the longitudinal Q-valley mass, and $\Delta_K$,
$\Delta_Q$ the spin-orbit splittings of the conduction band at K and Q. The direct gap at K is 1.71 (1.63) eV
without (with) spin-orbit coupling at the PAW geometry and changes by less than 0.06 eV over the other geometries.}
\label{tab:s2}
\small
\begin{tabular}{@{}lllllllll@{}}
\toprule
$a$ (\AA) & $z_S$ (\AA) & SOC & $E_{QK}$ (meV) & $q_Q$ & $m^*_K/m_0$ & $m^*_{Q,l}/m_0$ & $\Delta_K$ (meV) & $\Delta_Q$ (meV)\\
\midrule
3.181 & 1.576 (PAW) & no  & 190.0 & 0.510 & 0.48/0.50 & 0.59 & &\\
3.181 & 1.576 (PAW) & yes & 155.9 & 0.515 & 0.47/0.46 & 0.56 & 2.7 & 71.7\\
3.181 & 1.564 (NC)  & no  & 254.9 & 0.511 & 0.46/0.47 & 0.59 & &\\
3.181 & 1.564 (NC)  & yes & 219.7 & 0.515 & 0.45/0.44 & 0.58 & 2.9 & 75.2\\
3.160 & 1.565 (PAW fr.) & no  & 180.6 & 0.511 & 0.47/0.48 & 0.59 & &\\
3.160 & 1.565 (PAW fr.) & yes & 147.5 & 0.516 & 0.46/0.45 & 0.56 & 2.8 & 71.4\\
3.160 & 1.571 (NC)  & no  & 145.2 & 0.511 & 0.48/0.49 & 0.62 & &\\
3.160 & 1.571 (NC)  & yes & 112.6 & 0.515 & 0.48/0.46 & 0.61 & 2.7 & 70.6\\
\bottomrule
\end{tabular}
\end{table}

Four conclusions can be drawn from the simulations:

\begin{enumerate}
\item At a fixed geometry, the separation is independent of the pseudopotential family.
The norm-conserving bands at the PAW geometry reproduce the 190.0~meV of the PAW calculation,
its K-valley mass (0.48~$m_0$) and its band gap (1.71~eV).
\item The sulphur height is as important as the lattice constant. At fixed $a$, the
separation decreases by 5.7 to 5.8~eV per~\AA{} of sulphur height, that is by about
6~meV per 0.001~\AA. The two pseudopotential families relax the height to values
0.012~\AA{} apart at $a=3.181$~\AA{} (1.576 and 1.564~\AA), and this difference alone
moves the separation from 190 to 255~meV. The latter value agrees with the 246~meV that
Korm\'anyos~\etal{}~\cite{kormanyos2015} report for PBE, and the same effect explains the
two published LDA values of 90 and 200~meV at almost the same lattice
constant~\cite{kormanyos2013,kaasbjerg2012}.
\item For relaxed geometries, the separation decreases by about 165~meV per percent of
biaxial strain, so that the experimental lattice constant of 3.160~\AA{} gives 145~meV.
The measured pressure coefficient of $-65$~meV/GPa~\cite{fu2017} has the same sign.
\item Spin-orbit coupling lowers every value by 33 to 35~meV, one half of the Q-valley
splitting of 71 to 75~meV, whereas the K valley splits by only 3~meV. The valence-band
splitting at K of 146 to 148~meV agrees with the literature value of 148~meV~\cite{kormanyos2015}.
\end{enumerate}

At the experimental lattice constant with the relaxed sulphur height and spin-orbit
coupling, the separation is 113~meV, within 5~meV of the measured 118~meV. Because the
sulphur height of a monolayer is not known experimentally to the 0.005~\AA{} that would
fix the separation to 30~meV, this agreement places the measurement inside the range
spanned by the PBE geometries without singling out one geometry. Our transport model therefore uses the measured 118~meV, and the range from 70 to
255~meV is examined for its transport consequences below.

\subsection{Transport consequences}\label{sec:s1-emc}

\begin{figure}[ht]\centering
\includegraphics[width=0.75\textwidth]{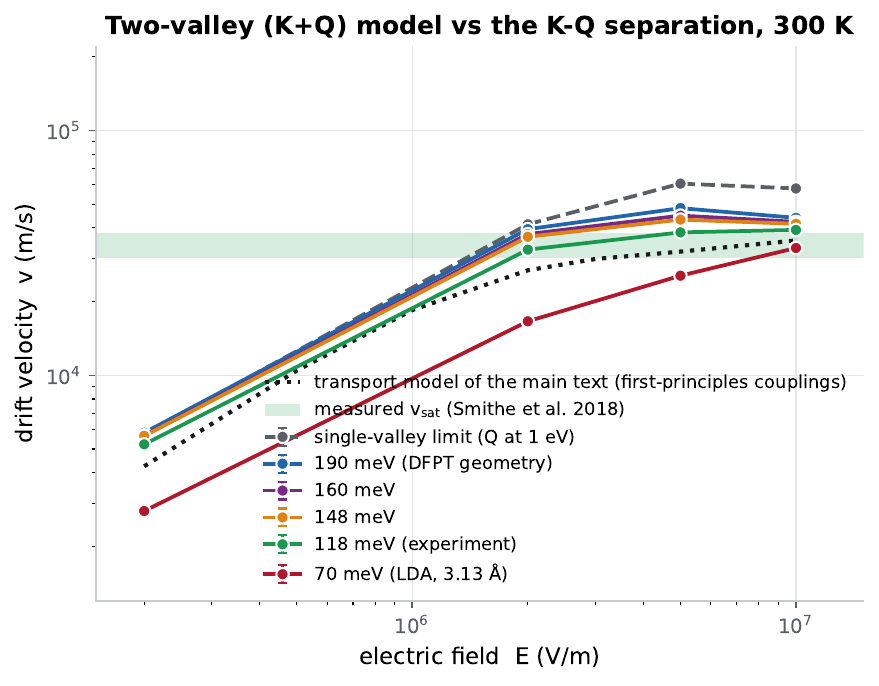}
\caption{Velocity-field characteristic of the two-valley model for the separations of
Table~\ref{tab:s3} and for its single-valley limit, together with the
transport model of the main text (K and the two spin-orbit-split Q sets,
first-principles couplings, four seeds) and the measured
saturation velocity of Smithe~\etal{}~\cite{smithe2018}. Error bars are standard errors over
seeds and are mostly smaller than the symbols.}
\label{fig:s2}
\end{figure}

To map the transport consequences of the separation over its published range, independently of our own transport model, we used the two-valley (K+Q) parameters of Pilotto~\etal{}~\cite{pilotto2022}, which is implemented in ViennaEMC alongside the
transport model of this work and shares its valley model. It describes the K valleys by an isotropic
non-parabolic band ($m^*=0.47\,m_0$, $\alpha=0.94$~eV$^{-1}$) and the six Q valleys by an
anisotropic non-parabolic band ($m_l=0.54\,m_0$, $m_t=1.14\,m_0$, $\alpha=1.16$~eV$^{-1}$)
placed $E_{QK}$ above K, with intravalley acoustic and zero-order intravalley and
intervalley optical deformation potentials for all K-K$'$, K-Q and Q-Q transitions from the
same reference. The separation sets the valley energies and, through the final-state
densities, every intervalley rate. We ran the suspended layer with intrinsic phonon
scattering only at 300~K for $E_{QK}=70$, 118, 148, 160 and 190~meV and, as the
single-valley limit of the same model, with the Q valleys placed at 1~eV. Each point is
the average over three seeds of 60\,000 steps of 0.1~fs after a 1.5~ps transient, and the
drift velocity is the occupation-weighted mean over the two valley types.
Table~\ref{tab:s3} and Fig.~\ref{fig:s2} show the results.

\begin{table}[ht]
\caption{Low-field mobility $\mu_0$ at 0.2~MV/m, drift velocity $v_{2}$, $v_{5}$ and
$v_{10}$ at 2, 5 and 10~MV/m (standard errors in parentheses) and Q-valley occupation
$p_Q$ at 0.2 and 10~MV/m of the two-valley (K+Q) transport model at 300~K for the suspended
monolayer with intrinsic phonons only, averaged over three seeds.}
\label{tab:s3}
\small
\begin{tabular}{@{}lllllll@{}}
\toprule
$E_{QK}$ (meV) & $\mu_0$ & $p_Q$ & $v_{2}$ & $v_{5}$ & $v_{10}$ & $p_Q$\\
 & (cm$^2$/Vs) & (0.2~MV/m) & ($10^4$~m/s) & ($10^4$~m/s) & ($10^4$~m/s) & (10~MV/m)\\
\midrule
single valley & 286 (3) & 0 & 4.14 (0.11) & 6.06 (0.04) & 5.79 (0.05) & 0\\
190 (PAW geometry) & 291 (4) & 0.003 & 3.95 (0.04) & 4.81 (0.04) & 4.40 (0.01) & 0.21\\
160 & 284 (4) & 0.008 & 3.77 (0.08) & 4.49 (0.05) & 4.23 (0.03) & 0.24\\
148 & 283 (3) & 0.013 & 3.68 (0.07) & 4.32 (0.03) & 4.16 (0.01) & 0.26\\
118 (experiment) & 261 (3) & 0.040 & 3.26 (0.03) & 3.83 (0.02) & 3.92 (0.01) & 0.30\\
70 (LDA, 3.13~\AA) & 140 (2) & 0.21 & 1.66 (0.03) & 2.55 (0.02) & 3.31 (0.01) & 0.41\\
\bottomrule
\end{tabular}
\end{table}

At the PBE separation of 190~meV, the Q valleys hold 0.3\,\% of the electrons at low field
and the mobility equals the single-valley limit of the same model within the statistical
error of 2\,\%. At the experimental 118~meV the occupation is 4\,\%, the mobility is 10\,\%
lower than at 190~meV, and the drift velocity at 2 and 5~MV/m is 17 and 20\,\% lower. At
70~meV the mobility halves, which reproduces the factor of two between
Li~\etal{}~\cite{li2013} and Kaasbjerg~\etal{}~\cite{kaasbjerg2012}. If the true separation were the 190~meV of the PAW geometry rather than the measured 118~meV used here, the low-field results of the main text would change by
10\,\% (Table~5 of the main text gives $+7$\,\% for this case) and the drift velocity at
5~MV/m by 20\,\%. Only a separation as low as the LDA value at a compressed lattice
constant, which is excluded by the measurement, would change the results by a factor of two.
The low-field results of the main text (temperature and substrate dependence, device and
sensor response, all at fields below 0.2~MV/m) therefore carry at most a 10\,\% uncertainty
from the valley model.

At high field, the Q valleys are populated for every separation. Above a few MV/m they
hold 21\,\% of the electrons at 10~MV/m even at 190~meV, and the drift velocity of the
two-valley model lies 21 to 24\,\% below the single-valley limit of the same model at
190~meV and 37\,\% below it at 118~meV. The two-valley model saturates at 3.9 to
4.8$\times10^{6}$~cm/s depending on the separation, consistent with the
$(3.4\pm0.4)\times10^{6}$~cm/s measured by Smithe~\etal{}~\cite{smithe2018}. Our transport model, with the first-principles couplings and the
spin-orbit-split Q valleys at 118~meV, saturates at $3.6\times10^{6}$~cm/s with a third
of the electrons in the Q valleys at 10~MV/m (Fig.~\ref{fig:s2}), inside the range obtained with the Pilotto parameters and in agreement with the measurement. The comparison of the separations is made with the Pilotto parameters throughout, because the two parametrizations also differ in their optical couplings and non-parabolicity (the single-valley limit of those parameters saturates
near $6\times10^{6}$~cm/s). The comparisons with the measured device in the main text are
made at drain fields of at most 0.1~MV/m over a 1~$\mu$m channel, in the linear part of
$v(E)$, and are not affected.

\subsection{Data}\label{sec:s1-data}

The Quantum ESPRESSO inputs and outputs of Table~\ref{tab:s2}, the wavevector list, the
extraction script, the ensemble Monte Carlo driver (\texttt{validation/mos2\_kq\_sensitivity.cpp}
in the ViennaEMC repository, with the compile-time override of the valley separation in the two-valley parameters) and the raw sweep data of Table~\ref{tab:s3} are part of the
dataset accompanying the main text.

\section{First-principles electron-phonon couplings}\label{sec:s2}

Our transport model uses electron-phonon couplings computed for this
work. This note documents the two independent DFPT calculations, gives the matrix elements by
phonon mode, wavevector and direction, states which values are used in the transport model and
why, and compares the two calculations with each other and with the reference parametrization of Kaasbjerg~\etal{}~\cite{kaasbjerg2012}. All calculations use Quantum
ESPRESSO~7.3.1~\cite{giannozzi2017} with the PBE functional~\cite{perdew1996}, the
norm-conserving scalar-relativistic PseudoDojo pseudopotentials~\cite{hamann2013,vansetten2018}
(14 valence electrons for Mo, 6 for S, so that the conduction band is band~14), a
plane-wave cutoff of 90~Ry (360~Ry for the density), a $12\times12\times1$ k-point mesh
and a hexagonal cell of height $6.33\,a$. The in-plane lattice constant and the sulphur
height were relaxed with the two-dimensional Coulomb cutoff~\cite{sohier2017} and the
cell free in the plane only, which gives $a=3.185$~\AA{} and $z_S=1.5625$~\AA{} (forces
below $10^{-4}$~Ry/bohr, in-plane stress below 0.1~kbar). At this geometry, the K-Q
separation without spin-orbit coupling is 265~meV, larger than at the PAW geometry of
Note~S1 because $a$ is larger and $z_S$ smaller, in line with the sensitivities
established there. The couplings are insensitive to this difference, and the valley
energy of the transport model is set from the measurement as described in Note~S1.

\subsection{Method}\label{sec:s2-method}

\begin{figure}[ht]\centering
\includegraphics[width=\textwidth]{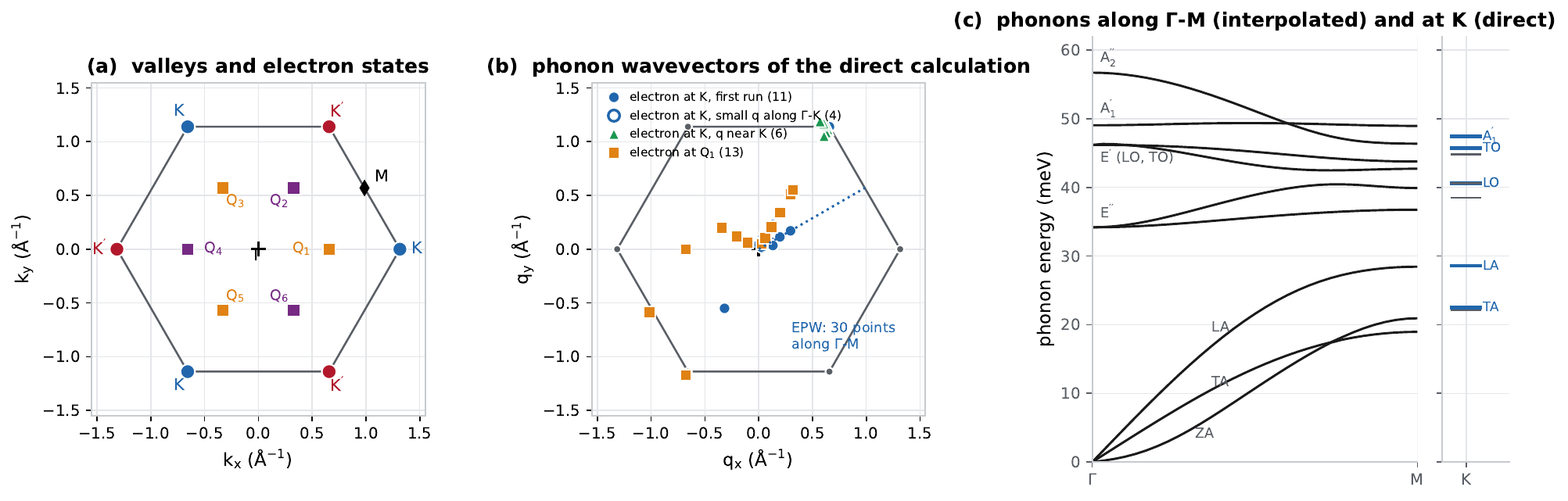}
\caption{(a) Brillouin zone of the monolayer with the K and K$'$ corners and the six Q
valleys ($\mathrm Q_1$ to $\mathrm Q_6$ at 0.5 of $\Gamma$-K, where the two colours mark the two
spin-orbit sets, $\mathrm Q_1$, $\mathrm Q_3$, $\mathrm Q_5$ and $\mathrm Q_2$, $\mathrm Q_4$,
$\mathrm Q_6$). The direct calculation places the electron at K and at $\mathrm Q_1$. (b) The
phonon wavevectors of the direct calculation, namely the eleven wavevectors of the first run with the
electron at K, the four small wavevectors along $\Gamma$-K added for the artefact
analysis, the six wavevectors near K for the first-order couplings, and the thirteen
wavevectors of the run with the electron at $\mathrm Q_1$. The dotted line is the
$\Gamma$-M segment on which the Wannier interpolation gives the couplings at 30 points.
(c) Phonon dispersion of the interpolation along $\Gamma$-M with the branch
labels used in the tables, and the nine modes at K from the dynamical matrix of the direct calculation, with the four modes of the K-K$'$ channels labelled at the energies
used in the transport model.}
\label{fig:s4}
\end{figure}

\paragraph{Direct DFPT matrix elements.} The DFPT matrix elements
$g_{\kappa\alpha}(\mathbf k,\mathbf q)=\langle\psi_{\mathbf k+\mathbf q}|\partial V/\partial u_{\kappa\alpha}(\mathbf q)|\psi_{\mathbf k}\rangle$
of the conduction band are written by the Allen-Heine-Cardona functionality of the phonon
code~\cite{lihm2020}, which stores them for a chosen initial state $\mathbf k$ and an
explicit phonon wavevector $\mathbf q$ without any interpolation. For each $\mathbf q$, the steps are a self-consistent calculation, a DFPT calculation of the dynamical
matrix and of the self-consistent potential derivative at $\mathbf q$, a
non-self-consistent calculation at $\mathbf k$ and $\mathbf k+\mathbf q$, and the
evaluation of the matrix elements. The implementation
requires norm-conserving pseudopotentials and does not support the two-dimensional Coulomb
cutoff, so this calculation relies on the vacuum of the cell (cell height 20.2~\AA) to isolate
the layer. This is adequate for the short-range and zone-boundary couplings but leaves the
long-range part of the acoustic and polar-optical couplings three-dimensional at
wavevectors below about $2\pi/L$, which is why the small-$q$ points are treated separately
below. The mode-resolved couplings follow from the projection of $g_{\kappa\alpha}$ on the
eigen-displacements $e^\nu_{\kappa\alpha}$ of the dynamical matrix,
$g_\nu=\sum_{\kappa\alpha}g_{\kappa\alpha}e^\nu_{\kappa\alpha}/\sqrt{2M_\kappa\omega_\nu}$,
and the deformation potentials are defined as in Ref.~\cite{kaasbjerg2012},
$\Xi_\nu=|g_\nu|\sqrt{2M_{\text{cell}}\omega_\nu}/|\mathbf q|$ for the acoustic and
$D_\nu=|g_\nu|\sqrt{2M_{\text{cell}}\omega_\nu}$ for the optical modes, with
$M_{\text{cell}}$ the mass of the unit cell. Three runs of this calculation were made, with the
electron at K and 15 wavevectors for the intravalley and K-K$'$ and K-Q couplings
(Table~\ref{tab:s4}), with the electron at K and six wavevectors within 0.1~\AA$^{-1}$ of
K for the first-order K-K$'$ couplings (Sec.~\ref{sec:s2-first}), and with the electron at
$\mathrm Q_1=0.515\,\Gamma$K and 13 wavevectors (Sec.~\ref{sec:s2-Q}).
Figure~\ref{fig:s4} shows the valleys, the sampled phonon wavevectors and the phonon
dispersion with the mode labels used throughout this note.

\paragraph{Wannier interpolation.} The second calculation uses the EPW
code~\cite{giustino2017,lee2023} with the two-dimensional Coulomb cutoff in the
self-consistent and DFPT steps and the two-dimensional dipole treatment of the long-range
electron-phonon coupling of Ponc\'e~\etal{}~\cite{ponce2023}. The dynamical matrices and
potential derivatives are computed on a $12\times12\times1$ q-mesh with the dielectric
tensor and Born charges, the electronic states on a $24\times24\times1$ k-mesh, and the
conduction and valence manifolds are represented by eleven maximally localised Wannier
functions~\cite{marzari2012} of Mo~$d$ and S~$p$ character (disentanglement window
$-12.18$ to $0.73$~eV, frozen window up to $0.8$~eV above the conduction-band minimum,
total spread 18.9~\AA$^2$, convergence $10^{-10}$~\AA$^2$). The interpolated bands
reproduce the DFT bands along $\Gamma$-M-K-$\Gamma$, and the couplings of the conduction
band at K are interpolated onto 30 wavevectors along $\Gamma$-M from 0.04 to
1.14~\AA$^{-1}$. It has the correct two-dimensional long-wavelength behaviour and is used to check the
direct calculation and to supply the polar-optical coupling, which is not accessible to the direct calculation. The same Wannier representation serves the phonon-limited
mobility calculation of Sec.~\ref{sec:s2-mob}.

\subsection{K-valley couplings}\label{sec:s2-K}

\begin{figure}[ht]\centering
\includegraphics[width=\textwidth]{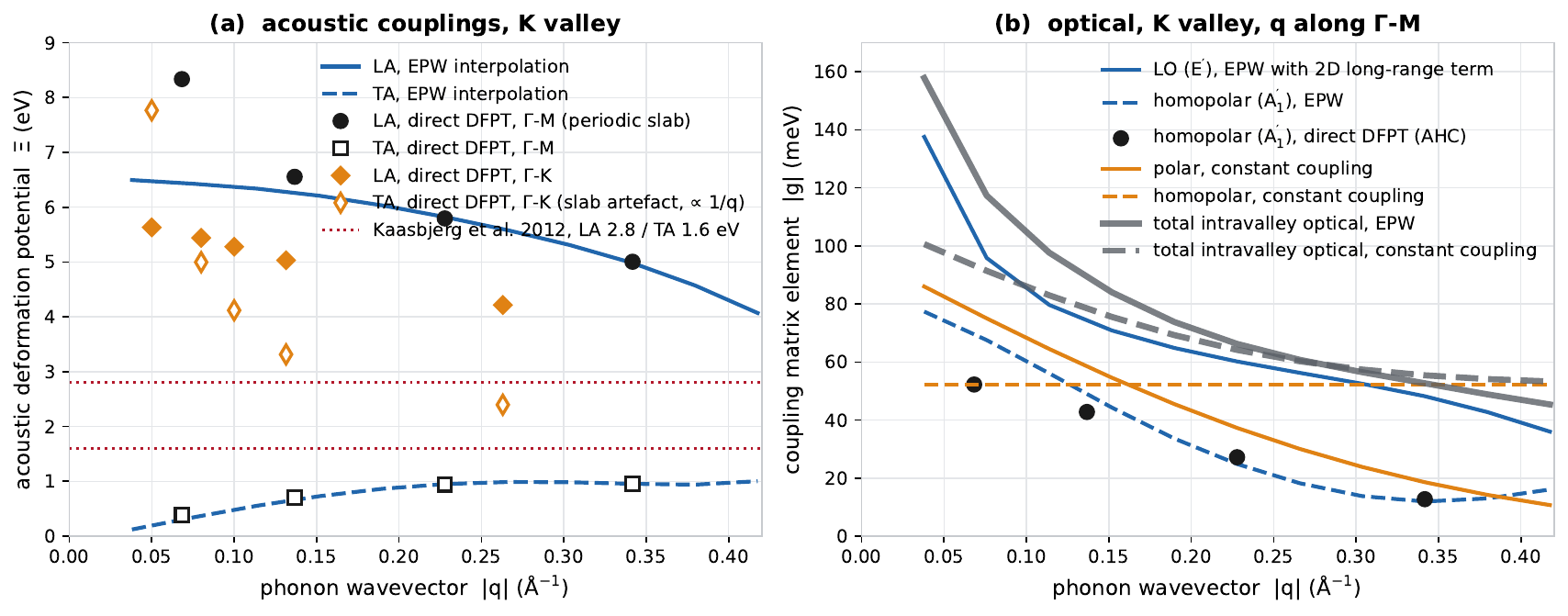}
\caption{Electron-phonon couplings of the K-valley conduction band. Lines and symbols
are the same DFPT matrix elements, interpolated through Wannier functions from a coarse
mesh (lines) and computed directly at each wavevector (symbols). (a) Acoustic deformation potentials from the Wannier interpolation with the
two-dimensional Coulomb cutoff along $\Gamma$-M (lines) and from the direct DFPT matrix
elements of the periodic slab along $\Gamma$-M (black symbols) and $\Gamma$-K (orange
symbols, Table~\ref{tab:s4}).
The $1/q$ rise of the direct LA coupling along $\Gamma$-M and of the direct TA coupling
along $\Gamma$-K is the long-range artefact of the slab without cutoff, whereas the direct
LA coupling along $\Gamma$-K is free of it and shows the genuine direction dependence.
The dotted lines are the reference values of Kaasbjerg~\etal{}~\cite{kaasbjerg2012}. (b) Optical coupling matrix elements, the LO
($E'$) and homopolar ($A_1'$) couplings of the interpolation and the direct calculation, the
polar and homopolar elements of the transport model, and the root-sum-square over the
three intravalley optical modes for the interpolation and for the transport model
(Table~\ref{tab:s5}).}
\label{fig:s3}
\end{figure}

Table~\ref{tab:s4} lists the couplings of the K-valley conduction band from the direct calculation and Fig.~\ref{fig:s3} compares them with the Wannier interpolation. Four observations determine the values used in the transport model.

\begin{table}[ht]
\caption{Electron-phonon couplings of the conduction band at K from the direct DFPT calculation by wavevector and direction. $\Xi_{\text{LA}}$ and $\Xi_{\text{TA}}$
are the acoustic deformation potentials, $D_{A_1'}$ the coupling to the homopolar
out-of-plane optical mode (11.9~THz, 49~meV) and $D_{E'}$ the root-sum-square coupling to
the two in-plane optical modes (11.2~THz, 46~meV), which in this calculation contains the
three-dimensional long-range part, so that the transport model uses the interpolated value of
Table~\ref{tab:s5} instead. The 0.034~\AA$^{-1}$ point lies inside the acoustic-sum-rule
residual of the dynamical matrix and is not used. The rise of $\Xi_{\text{LA}}$ along
$\Gamma$-M and of $\Xi_{\text{TA}}$ along $\Gamma$-K at small wavevector is the
periodic-slab artefact discussed in the text. The last two rows are the intervalley
transitions with the electron at K, listed by phonon mode with its energy.}
\label{tab:s4}
\small
\begin{tabular}{@{}llllll@{}}
\toprule
$|\mathbf q|$ (\AA$^{-1}$) & direction & $\Xi_{\text{LA}}$ (eV) & $\Xi_{\text{TA}}$ (eV) & $D_{A_1'}$ (eV/\AA) & $D_{E'}$ (eV/\AA)\\
\midrule
0.050 & $\Gamma$-K & 5.63 & 7.77 & 3.32 & 8.44\\
0.068 & $\Gamma$-M & 8.3 & 0.39 & 3.21 & 6.58\\
0.080 & $\Gamma$-K & 5.44 & 5.00 & 3.14 & 5.88\\
0.080 & $15^\circ$ from $\Gamma$-M & 6.62 & 3.52 & 3.13 & 5.88\\
0.100 & $\Gamma$-K & 5.28 & 4.12 & 2.99 & 5.12\\
0.132 & $\Gamma$-K & 5.03 & 3.31 & 2.72 & 4.45\\
0.137 & $\Gamma$-M & 6.55 & 0.71 & 2.63 & 4.40\\
0.137 & $15^\circ$ from $\Gamma$-M & 6.56 & 2.32 & 2.59 & 4.41\\
0.137 & $\Gamma$-K & 5.00 & 3.22 & 2.67 & 4.40\\
0.228 & $\Gamma$-M & 5.79 & 0.94 & 1.67 & 3.71\\
0.263 & $\Gamma$-K & 4.22 & 2.40 & 1.43 & 3.64\\
0.342 & $\Gamma$-M & 5.00 & 0.96 & 0.78 & 3.07\\
\midrule
\multicolumn{6}{@{}l@{}}{K$\to$K$'$ ($\mathbf q=$ K, 1.315~\AA$^{-1}$): $E'$ mode at 9.8~THz (40.6~meV), $D=1.66$~eV/\AA;}\\
\multicolumn{6}{@{}l@{}}{\quad all other modes below 0.1~eV/\AA}\\
\multicolumn{6}{@{}l@{}}{K$\to$Q ($\mathbf q=$ Q$-$K, 0.635~\AA$^{-1}$): LA 22.2~meV, 1.15; LO 42.7~meV, 1.44; $A_1'$ 49.3~meV, 1.04;}\\
\multicolumn{6}{@{}l@{}}{\quad TA 45.5~meV, 0.57; TA 16.5~meV, 0.22~eV/\AA}\\
\bottomrule
\end{tabular}
\end{table}

\begin{enumerate}
\item The direct calculation carries a long-range artefact at small wavevector. Along
$\Gamma$-K its TA coupling is 7.8, 5.0, 4.1 and 3.3~eV at 0.05, 0.08, 0.10 and
0.13~\AA$^{-1}$, that is $\Xi_{\text{TA}}\,q=0.39$ to 0.43~eV/\AA$^{-1}$, a $1/q$ law that cannot arise from a deformation potential, and along $\Gamma$-M the same term appears in the LA
channel (11.1 and 8.3~eV at 0.034 and 0.068~\AA$^{-1}$). This is the piezoelectric-like
coupling of the periodic slab without Coulomb cutoff, three-dimensional at wavevectors
below about $2\pi/L$ and with the three-fold angular pattern of the hexagonal layer
(longitudinal along $\Gamma$-M, transverse along $\Gamma$-K). The interpolation with the cutoff is free of this term, as its LA coupling tends smoothly to 6.5~eV and
its TA coupling to zero at $q\to0$, as required for a non-degenerate band, and the
two-dimensional piezoelectric coupling that it does contain is small. An angle average
over the direct calculation at 0.137~\AA$^{-1}$ would give $\Xi_{\text{TA}}=2.3$~eV and
overstate the TA coupling about three-fold. The transport model uses only the artefact-free data.
\item The genuine couplings are direction dependent at finite wavevector. The LA
coupling along $\Gamma$-K, free of the artefact by symmetry, is 5.6, 5.4, 5.3, 5.0 and
4.2~eV at 0.05, 0.08, 0.10, 0.13 and 0.26~\AA$^{-1}$ against 6.5, 6.4, 6.3, 6.2 and
5.8~eV along $\Gamma$-M from the interpolation, and the TA coupling after subtracting the
$1/q$ term in quadrature is about 1.0~eV at 0.13 and 1.8~eV at 0.26~\AA$^{-1}$ along
$\Gamma$-K and at $15^\circ$, against 0.7 and 0.95~eV along $\Gamma$-M. The tables of Sec.~\ref{sec:s2-kernel} are the interpolated $\Gamma$-M curves multiplied
by the root-mean-square ratio over directions, 1.0 below 0.05~\AA$^{-1}$ (isotropic
limit), 0.91 (LA) and 1.3 (TA) at 0.137 and 0.90 and 1.5 at 0.30~\AA$^{-1}$.
\item The LA coupling is about twice the reference value of 2.8~eV~\cite{kaasbjerg2012},
which was obtained from an LDA calculation with a different projection, and the TA
coupling is below the reference 1.6~eV. The LA coupling is therefore the quantity in which the two first-principles calculations differ most, and the transport model uses the computed
wavevector-dependent values without any adjustable parameter.
\item The homopolar coupling decays with wavevector in both calculations, from 4.7 (interpolation)
or 3.4~eV/\AA{} (direct) at the smallest wavevectors to 2.7~eV/\AA{} at 0.15~\AA$^{-1}$
and 0.8~eV/\AA{} at 0.30~\AA$^{-1}$, where the two calculations agree within 10\,\%. The transport model
uses the interpolated $A_1'$ matrix element as a table (Sec.~\ref{sec:s2-polar}). The K-K$'$
intervalley coupling through the $E'$ mode at K, $1.7\times10^{8}$~eV/cm, is 0.65 of the
reference value. The out-of-plane $A_2''$ and the $E''$ modes couple weakly or not at all
(below 0.4 and 0.1~eV/\AA), consistent with the mirror symmetry of the layer, and the
first-order K-K$'$ couplings, which vanish for $\mathbf q=$ K exactly, are obtained from
wavevectors near K (Sec.~\ref{sec:s2-first}).
\end{enumerate}

\subsection{Polar-optical coupling and the optical sum}\label{sec:s2-polar}

The polar coupling to the LO branch of the $E'$ mode can only be obtained from the calculation with the two-dimensional treatment. Table~\ref{tab:s5} and Fig.~\ref{fig:s3}(b) give the
interpolated LO, TO and $A_1'$ matrix elements, which the transport model uses directly, together
with a constant-coupling representation that quantifies the effect of the wavevector
dependence, namely the two-dimensional Fr\"ohlich element of Kaasbjerg~\etal{}~\cite{kaasbjerg2012},
$g(q)=g_0\,\mathrm{erfc}(\sigma q/2)$ with $g_0=C/\sqrt{A_{\text{cell}}}=97$~meV
($C=0.286$~eV\,\AA, $A_{\text{cell}}=8.65$~\AA$^2$) and $\sigma=5.41$~\AA, and the constant
homopolar element $|g|=52$~meV that corresponds to $D_{A_1'}=3.2\times10^{8}$~eV/cm.
The interpolated LO coupling is finite at $q\to0$, as required in two dimensions, and is
reduced with increasing wavevector by the dielectric screening of the layer itself. The
strictly two-dimensional limit follows analytically from the same DFPT
run~\cite{sohier2016}. Our in-plane Born charges ($Z^*_{\text{Mo}}=-0.99$,
$Z^*_{\text{S}}=0.49$), the $E'$ eigenvector and the zero-point displacement of the Mo-S
reduced mass give $g_0=e\,p_{\text{zp}}/(2\varepsilon_0A_{\text{cell}})=0.35$~eV, which the
layer screens as $g_0/(1+r_0q)$ with $r_0=2\pi\alpha_{2D}=43.5$~\AA{} from our in-plane
dielectric constant. This form gives 132~meV at 0.038~\AA$^{-1}$ against the 138~meV of
the interpolation, which confirms the long-range treatment. At larger wavevectors, the interpolated coupling
decays more slowly than the simple screening form (81 against 96~meV at 0.076~\AA$^{-1}$)
because the interpolation carries the full wavevector dependence of the slab response.

\begin{table}[ht]
\caption{Intravalley optical coupling matrix elements of the conduction band at K along
$\Gamma$-M from the Wannier interpolation with the two-dimensional long-range treatment
(LO and TO branches of the $E'$ mode, 46~meV, and the $A_1'$ mode, 49~meV), the two
elements of the constant-coupling representation described in the text, and the root-sum-square (rss) of the intravalley
optical couplings of both, in meV. The two $E'$ branches cross at 0.15~\AA$^{-1}$ and are
identified by their coupling.}
\label{tab:s5}
\begin{tabular*}{\textwidth}{@{\extracolsep{\fill}}lllllllll@{}}
\toprule
$|\mathbf q|$ (\AA$^{-1}$) & LO & TO & $A_1'$ & rss & constant polar & constant $A_1'$ & rss & ratio\\
\midrule
0.038 & 138 & 2 & 77 & 158 & 86 & 52 & 101 & 1.57\\
0.076 & 96 & 5 & 68 & 117 & 75 & 52 & 91 & 1.28\\
0.114 & 80 & 7 & 56 & 98 & 65 & 52 & 83 & 1.18\\
0.152 & 71 & 9 & 44 & 84 & 55 & 52 & 76 & 1.11\\
0.228 & 60 & 12 & 25 & 66 & 37 & 52 & 64 & 1.03\\
0.304 & 53 & 16 & 14 & 57 & 24 & 52 & 57 & 0.98\\
\bottomrule
\end{tabular*}
\end{table}

The constant polar element lies a factor of 1.2 to 1.3 below the interpolated LO
coupling in the thermal window and a factor of two below it at 0.3~\AA$^{-1}$, whereas the
constant homopolar element exceeds the interpolated value for all wavevectors
above 0.07~\AA$^{-1}$. All three intravalley optical modes lie within 3~meV of each other
(46 to 49~meV) and affect the transport through the same emission threshold and the same
final-state density, so their combined strength is the relevant quantity. Taken as
root-sum-squares, the three interpolated couplings and the two constant elements
agree within 20\,\% for $|\mathbf q|\ge0.1$~\AA$^{-1}$ and within 3\,\% above
0.2~\AA$^{-1}$. Their difference reaches 28\,\% at 0.08 and 57\,\% at 0.04~\AA$^{-1}$, where
the phase space for a thermal electron is small. A constant-coupling representation
therefore reproduces the total intravalley optical strength within 10 to 20\,\%, with a
different division between the polar and the homopolar channel than the interpolation.
The transport model avoids this difference by using the interpolated $|g(q)|$ of both modes
directly, with the free-carrier screening applied to the polar channel only.

\subsection{First-order K-K$'$ couplings}\label{sec:s2-first}

The first-order intervalley couplings are linear in the phonon wavevector measured
from K, with
$|g|\propto|\mathbf q-\mathrm K|=|\boldsymbol\delta'-\boldsymbol\delta|$ for
an electron at $\mathrm K+\boldsymbol\delta$ scattered to $\mathrm K'+\boldsymbol\delta'$.
They therefore vanish for $\mathbf q=\mathrm K$ irrespective of the electron position,
and we obtain them from matrix elements with the electron at K and the phonon at
$\mathrm K+\delta\mathbf q$. Table~\ref{tab:s5b} lists $D_{\text{eff}}/|\delta\mathbf q|$
for $|\delta\mathbf q|=0.03$, 0.06 and 0.10~\AA$^{-1}$ along K-$\Gamma$ and
perpendicular to it. The TA, TO and $A_1'$-like couplings are linear in
$|\delta\mathbf q|$ and isotropic to within 10\,\%, so each is a well-defined
first-order deformation potential in the sense of Ref.~\cite{kaasbjerg2012}, whose
angular average $|\boldsymbol\delta|^2+|\boldsymbol\delta'|^2$ is used by the transport model. For the weaker and less regular LA coupling we use the mean value. The zero-order coupling
of the LO-like $E'$ mode is recovered unchanged at every $\delta\mathbf q$
(1.65~eV/\AA), which confirms the separation into zero- and first-order channels.
Compared with the reference parametrization, the TA and TO first-order couplings are two to three
times larger (four and six times in the rate) and a fourth channel, the $A_1'$-like
mode at K, is present.

\begin{table}[ht]
\caption{First-order K-K$'$ intervalley couplings $\Xi_1=D_{\text{eff}}/|\delta\mathbf q|$
(eV) of the conduction band from the direct calculation with the electron at K and the
phonon at $\mathrm K+\delta\mathbf q$, for three magnitudes of $\delta\mathbf q$ along
K-$\Gamma$ and perpendicular to it; the last columns give the value used in the transport model and
the reference value~\cite{kaasbjerg2012}. The flexural, $E''$ and $A_2''$ modes give
zero.}
\label{tab:s5b}
\small
\begin{tabular}{@{}llllll@{}}
\toprule
Mode at K & $\hbar\omega$ (meV) & K-$\Gamma$ (0.03 / 0.06 / 0.10~\AA$^{-1}$) & perpendicular & transport model & reference\\
\midrule
TA & 22.5 & 12.3 / 12.1 / 11.6 & 12.3 / 11.9 / 11.2 & 12.0 & 5.9\\
LA & 28.5 & 1.6 / 1.3 / 0.9 & 2.1 / 2.4 / 2.8 & 2.0 & 3.9\\
LO ($E'$) & 40.6 & zero order, 1.65~eV/\AA & same & $1.7\times10^{8}$~eV/cm & $2.6\times10^{8}$~eV/cm\\
TO ($E'$) & 45.7 & 4.7 / 4.9 / 5.2 & 4.6 / 4.6 / 4.7 & 4.8 & 1.9\\
$A_1'$-like & 47.4 & 3.7 / 3.4 / 2.9 & 3.9 / 3.8 / 3.7 & 3.5 & not included\\
\bottomrule
\end{tabular}
\end{table}

\subsection{Q-valley couplings}\label{sec:s2-Q}

The second run of the direct calculation places the electron at $\mathrm Q_1$ and covers the
intravalley couplings for longitudinal and transverse wavevectors (Table~\ref{tab:s6}) and
the intervalley transitions to the five other Q valleys, to K$'$ and back to K
(Table~\ref{tab:s7}). Three points are specific to the Q valley. The intravalley acoustic
couplings are smaller than at K and less anisotropic (2.7 and 3.2~eV for the longitudinal
mode with longitudinal and transverse wavevector, 3.4 and 0.3~eV for the transverse mode),
so that the transport model uses $\Xi_L=3.0$ and $\Xi_T=2.4$~eV, the averages at
0.12~\AA$^{-1}$. The flexural (ZA) mode, which does not couple at K because of the
mirror symmetry of the layer, couples to the Q valley for transverse wavevectors
(2.6~eV equivalent at 0.12~\AA$^{-1}$, 0.8~eV at 0.24~\AA$^{-1}$). This channel is not
part of the transport model and would add at most a few per cent to the acoustic rate of the Q
valleys. The homopolar coupling at Q is larger than at K, $5.4\times10^{8}$~eV/cm
below 0.12~\AA$^{-1}$, and decays with wavevector as at K. As a consistency check, the
Q$\to$K matrix elements at $\mathbf q=$ K$-$Q$_1$ reproduce the K$\to$Q elements of the
first run at $\mathbf q=$ Q$-$K (Table~\ref{tab:s4}) within 1\,\% for all five phonon
channels, as time-reversal symmetry requires.

\begin{table}[ht]
\caption{Intravalley couplings of the Q-valley conduction band (electron at
$\mathrm Q_1=0.515\,\Gamma$K) from the direct calculation for wavevectors along the valley axis
(longitudinal, towards $\Gamma$) and perpendicular to it (transverse). $\Xi_L$ and $\Xi_T$
refer to the longitudinal and transverse acoustic modes, $\Xi_{\text{ZA}}$ to the
flexural mode in the same units, $D_{A_1'}$ to the homopolar mode.}
\label{tab:s6}
\small
\begin{tabular}{@{}llllll@{}}
\toprule
$|\mathbf q|$ (\AA$^{-1}$) & direction & $\Xi_L$ (eV) & $\Xi_T$ (eV) & $\Xi_{\text{ZA}}$ (eV) & $D_{A_1'}$ (eV/\AA)\\
\midrule
0.059 & longitudinal & 2.63 & 6.43 & 0 & 5.53\\
0.118 & longitudinal & 2.70 & 3.37 & 0 & 5.36\\
0.118 & transverse & 3.22 & 0.29 & 2.60 & 5.51\\
0.237 & longitudinal & 2.95 & 1.78 & 0 & 4.25\\
0.237 & transverse & 1.24 & 0.59 & 0.84 & 4.30\\
0.395 & longitudinal & 2.65 & 0.53 & 0 & 1.43\\
0.395 & transverse & 2.50 & 1.58 & 0.06 & 1.72\\
0.592 & longitudinal & 2.05 & 0.28 & 0 & 0.69\\
\bottomrule
\end{tabular}
\end{table}

\begin{table}[ht]
\caption{Intervalley couplings with the electron at $\mathrm Q_1$ from the direct calculation,
by phonon channel (mode energy in meV, zero-order deformation potential in
$10^{8}$~eV/cm). Modes with couplings below $0.5\times10^{8}$~eV/cm are omitted. The
Q$\to$K row is the reverse of the K$\to$Q transition of Table~\ref{tab:s4}. The K$\to$Q$'$
transitions to the three far Q valleys use the Q$\to$K$'$ elements by time reversal.}
\label{tab:s7}
\small
\begin{tabular*}{\textwidth}{@{\extracolsep{\fill}}llp{6.0cm}l@{}}
\toprule
transition & $|\mathbf q|$ (\AA$^{-1}$) & channels (meV, $10^{8}$~eV/cm) & used for\\
\midrule
Q$_1\to$Q$_2$, Q$_6$ ($60^\circ$) & 0.677 & LA 23.0, 3.87; $A_1'$ 49.2, 2.07; TO 45.4, 1.41; TA 17.4, 0.78; LO 42.4, 0.74 & Q$\to$Q nearest\\
Q$_1\to$Q$_3$, Q$_5$ ($120^\circ$) & 1.173 & LO 42.8, 3.15; $A_1'$ 49.1, 2.97; LA 28.5, 1.93 & Q$\to$Q next-nearest\\
Q$_1\to$Q$_4$ ($180^\circ$) & 1.355 & 28.5, 5.09; 22.5, 2.92; 47.4, 2.43; 40.6, 1.44 & Q$\to$Q opposite\\
Q$_1\to$K$'$ & 1.992 & $A_1'$ 49.1, 4.95; LA 28.5, 4.64; LO 43.9, 2.86; 19.0, 1.56 & Q$\to$K$'$, K$\to$Q$'$ (far)\\
Q$_1\to$K & 0.638 & LA 22.2, 1.15; LO 42.7, 1.44; $A_1'$ 49.3, 1.03; TA 45.5, 0.58; TA 16.5, 0.22 & Q$\to$K, K$\to$Q (near)\\
\bottomrule
\end{tabular*}
\end{table}

\subsection{Values used in the transport model}\label{sec:s2-kernel}

\begin{figure}[ht]\centering
\includegraphics[width=\textwidth]{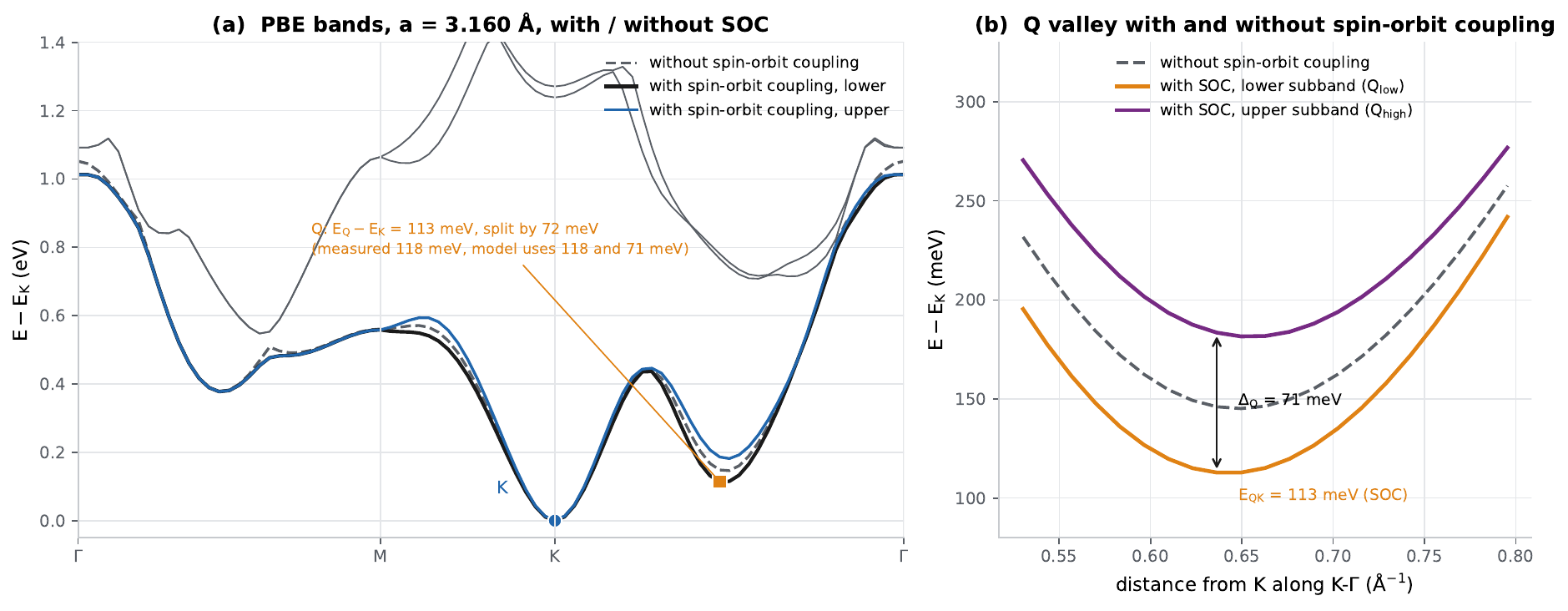}
\caption{(a) Lowest conduction bands of PBE along $\Gamma$-M-K-$\Gamma$ at the
experimental lattice constant with the relaxed sulphur height (the geometry of Note~S1
closest to experiment), with and without spin-orbit coupling. The K valley splits by
3~meV and the Q valley by 72~meV, and the lower Q subband lies 113~meV above K against
the measured 118~meV. The transport model uses the measured separation with this splitting. (b)
The Q valley along K-$\Gamma$ from the same calculations, the scalar-relativistic band
and the two spin-orbit subbands that the transport model represents as the sets
$\mathrm Q_{\text{low}}$ and $\mathrm Q_{\text{high}}$. The isotropic Kane band of the K
valley used in the transport model ($m^*=0.446\,m_0$, $\alpha=1.16$~eV$^{-1}$) is fitted to the
Wannier-interpolated band of the electron-phonon calculations, and is not changed by the 3~meV
splitting at K.}
\label{fig:s5}
\end{figure}

For the K valley, the transport model uses the isotropic Kane band with
$m^*=0.446\,m_0$ and $\alpha=1.16$~eV$^{-1}$, fitted to the interpolated dispersion
between 25 and 150~meV in both directions. Acoustic scattering follows the tables
$\Xi_{\text{LA}}(q)$ and $\Xi_{\text{TA}}(q)$ of Sec.~\ref{sec:s2-K} (6.5, 5.65 and
4.8~eV and 0, 0.95 and 1.45~eV at 0, 0.137 and 0.30~\AA$^{-1}$), and polar and homopolar
scattering the LO and $A_1'$ tables of Table~\ref{tab:s5}, joined below
0.04~\AA$^{-1}$ to the analytic limit of Sec.~\ref{sec:s2-polar}. Intervalley
scattering uses the zero-order K-K$'$ coupling of $1.7\times10^{8}$~eV/cm at 40.6~meV
and the first-order couplings of Table~\ref{tab:s5b}, with an intravalley
TO($\Gamma$) coupling of 3.6~eV from the interpolated TO branch. Mass density, sound
velocities and phonon energies are those of the computed lattice
($3.03\times10^{-6}$~kg/m$^2$, 6580 and 4020~m/s, 49.2, 46.3 and 40.6~meV). For the Q valleys it uses the band parameters of Note~S1 ($m_l=0.646$, $m_t=1.258$,
$\alpha=0.3$~eV$^{-1}$), the measured separation $E_{QK}=118$~meV, the computed
spin-orbit splitting of 71~meV (Fig.~\ref{fig:s5}), the intravalley values $\Xi_L=3.0$
and $\Xi_T=2.4$~eV with the $A_1'$ table of Table~\ref{tab:s6}, and the intervalley
channels of Table~\ref{tab:s7} together with the K$\to$Q channels of
Table~\ref{tab:s4}. Each channel is a separate zero-order mechanism with its own phonon
energy, and the sub-valley assignment follows from the spin selection (Sec.~2.2 of the
main text). There is no piezoelectric term. The free-carrier screening
of the polar coupling and the extrinsic mechanisms (remote substrate phonons, charged
impurities, roughness) are those of the main text and act on all three valley types. No
number in this list is taken from another calculation.

\subsection{Phonon-limited mobility from the Boltzmann equation}\label{sec:s2-mob}

The same Wannier representation is used to solve the linearised Boltzmann equation for
the phonon-limited electron mobility of the suspended monolayer between 150 and 400~K
(EPW~\cite{ponce2016,lee2023}), which checks the ensemble Monte Carlo transport model by a method
that shares the first-principles couplings but none of its approximations. The
Boltzmann solution carries the full wavevector and direction dependence of every matrix
element, the trigonally warped band and the exact angular structure of the scattering,
whereas the transport model uses the isotropic tables of Sec.~\ref{sec:s2-kernel}, the Kane band
and analytic deflection-angle distributions. Electrons and phonons were interpolated on
$200\times200$ meshes at a carrier density of $10^{12}$~cm$^{-2}$, at which the electron
gas is non-degenerate at every temperature considered, so that the mobility summed over
all states coincides with the mobility over the states above the Fermi level that the
code reports by default. Both the relaxation-time (SERTA) and the iterative (IBTE)
solutions were obtained. The
interpolation uses the DFPT geometry without spin-orbit coupling, so the Q valleys lie
265~meV above K and are empty at these temperatures. For a comparison under identical conditions, the
transport model was run with the Q valleys at the same 265~meV and with the free-carrier screening
of the polar coupling switched off (eight seeds, 0.2~MV/m), because the Boltzmann
calculation is for the undoped layer.

\begin{figure}[ht]\centering
\includegraphics[width=\textwidth]{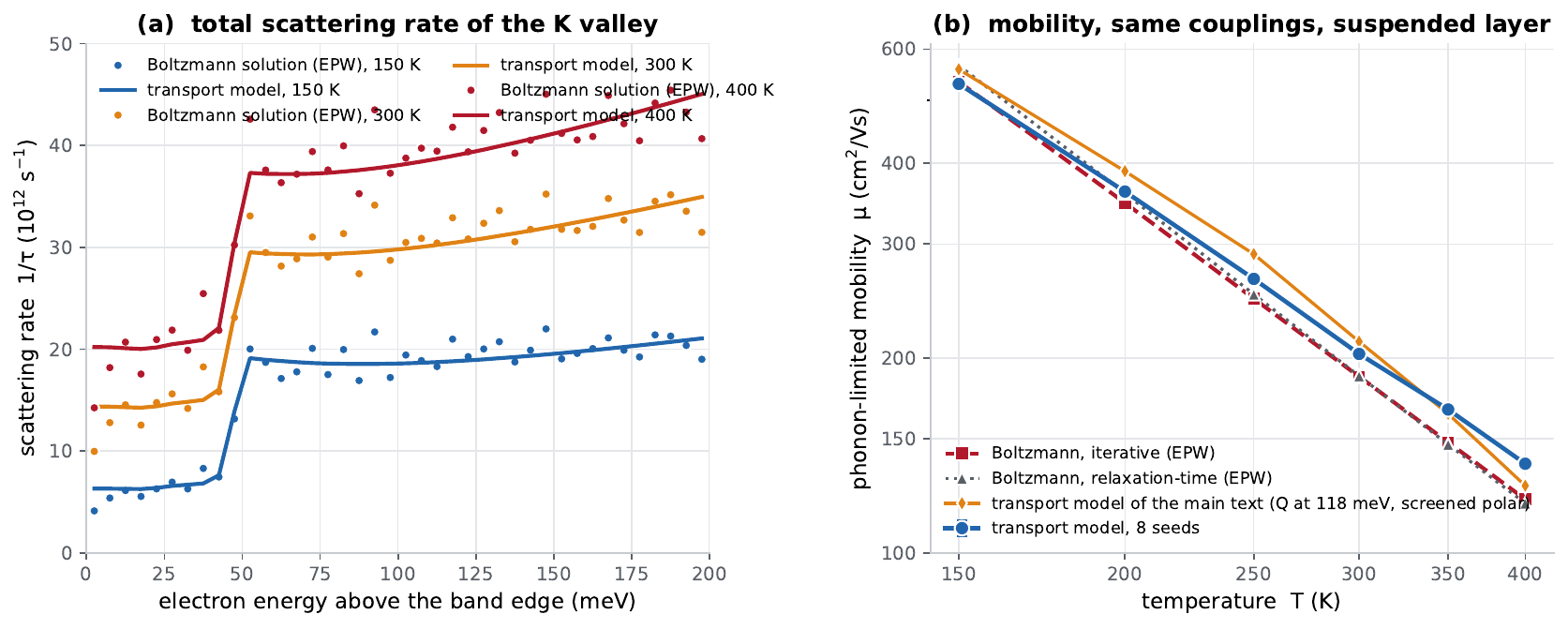}
\caption{The transport model against the Boltzmann solution of the Wannier interpolation. (a) Total
scattering rate of a K-valley electron versus its energy above the band edge at 150, 300
and 400~K, the state-resolved rates of the Boltzmann solution averaged in 5~meV bins
(symbols) and the rates of the transport model with the same couplings (lines). (b)
Phonon-limited mobility of the suspended layer, the transport model with the Q valleys at 265~meV
and the polar coupling unscreened (circles, standard errors smaller than the symbols),
the iterative (squares) and relaxation-time (triangles) Boltzmann solutions, and the
transport model of the main text with the Q valleys at the measured 118~meV and the polar coupling
screened by the free carriers (diamonds).}
\label{fig:s6}
\end{figure}

Figure~\ref{fig:s6}(a) compares the total scattering rate of a K-valley electron as a
function of its energy. Above 10~meV the two agree within 5\,\% at all three
temperatures, below and above the optical emission threshold at 46 to 49~meV, where the
rate rises by a factor of 2.7 at 150~K through the onset of $A_1'$ and LO emission. The
only larger deviations are single 5~meV bins at 15 to 20~meV, where the Boltzmann rates
scatter, and at the threshold, where the onset of the transport model is sharper than that of
the interpolated rates (14 and 12\,\%). In the first 10~meV above the band edge, the rate of the transport model exceeds the
Boltzmann rate, by 60 to 80\,\% in the first 2.5~meV and by 10 to 20\,\% between 2.5 and
10~meV. The elastic equipartition form of the acoustic channels neglects the phonon
energy, whereas in the Boltzmann solution an electron within a few meV of the band edge
cannot emit an acoustic phonon of 1 to 4~meV.
These states carry 18\,\% of the energy-weighted mobility sum at 150~K and 6\,\% at
300~K, so the excess lowers the mobility of the transport model by about 3\,\% at 150~K and 1\,\% at
300~K.

\begin{table}[ht]
\caption{Phonon-limited mobility of the suspended monolayer (cm$^2$/Vs) from the
Boltzmann solution of the Wannier interpolation (SERTA and IBTE, $200\times200$ meshes,
$10^{12}$~cm$^{-2}$, Q valleys at 265~meV, no free-carrier screening) and from the
transport model under the same conditions (eight seeds, standard errors in parentheses), and the
transport model of the main text (Q valleys at the measured 118~meV with the spin-orbit splitting, polar
coupling screened at $10^{12}$~cm$^{-2}$). The last row gives the exponent of a power-law
fit over 150 to 400~K.}
\label{tab:s8}
\small
\begin{tabular}{@{}lcccc@{}}
\toprule
$T$ (K) & Boltzmann, SERTA & Boltzmann, IBTE & EMC, same conditions & EMC, main text\\
\midrule
150 & 566 & 534 & 530 (2) & 558 (2)\\
200 & 359 & 347 & 362 (2) & 389 (3)\\
250 & 251 & 247 & 265 (2) & 289 (2)\\
300 & 188 & 187 & 203 (2) & 212 (1)\\
350 & 147 & 148 & 167 (2) & 164 (2)\\
400 & 119 & 121 & 137 (1) & 127 (3)\\
\midrule
$\gamma$ & 1.59 & 1.52 & 1.38 & 1.51\\
\bottomrule
\end{tabular}
\end{table}

Table~\ref{tab:s8} and Fig.~\ref{fig:s6}(b) compare the mobilities. Under the same
conditions, the transport model agrees with the iterative Boltzmann solution within 1\,\% at 150~K
and 4\,\% at 200~K, and exceeds it by 7, 8, 12 and 13\,\% at 250, 300, 350 and 400~K.
The difference grows with temperature although the total rates agree at every energy, so
it does not come from the couplings but from the two approximations of the transport model that are not made in the Boltzmann solution. The deflection angle after each event is sampled from
the isotropic average of $|g(q)|^2$ over the direction of $\mathbf q$, and the K valley is
an isotropic Kane band, whereas the trigonal warping of the real band grows above
150~meV, where an increasing fraction of the electrons resides at 350 and 400~K. Both
approximations affect the momentum relaxation rather than the total rate, and the
difference between the relaxation-time and the iterative Boltzmann solutions (6\,\% at
150~K, 1\,\% at 400~K) shows the size of such angular effects in this material. We
therefore take 13\,\% at 400~K, and 8\,\% at room temperature, as the bound of these approximations on the phonon-limited mobility.

Our transport model differs from this run in two physical inputs. The Q
valleys at the measured 118~meV, with the spin-orbit splitting of Note~S1, add
intervalley scattering and lower the mobility by 6\,\% at 300~K and 21\,\% at 400~K
(exponent 1.38 to 1.60). Static free-carrier screening of the polar coupling at
$10^{12}$~cm$^{-2}$ raises it by 5\,\% at 150~K, 11\,\% at 300~K and 17\,\% at 400~K
(exponent 1.60 to 1.51). We apply this screening because the plasma energy of the sheet
at the relevant wavevectors (84 to 145~meV for $q$ between 0.05 and 0.15~\AA$^{-1}$ at
$10^{12}$~cm$^{-2}$) exceeds the LO phonon energy of 46~meV, so that the free carriers
follow the polar phonon field. The static form overestimates the screening by at most
30\,\% where the two energies are closest, which bounds its effect on the
mobility used to obtain the results presented in the main text at about 4\,\% at 300~K.

\subsection{Data}\label{sec:s2-data}

The inputs and outputs of both calculations (the explicit wavevector lists, the DFPT and
matrix-element outputs of the 34 wavevectors, the Wannier interpolation output and the
extraction scripts), the parameter header of the transport model
(\texttt{examples/singleLayerMoS2/parameterTwoValley.hpp} in the ViennaEMC repository,
with the reference parametrization of Ref.~\cite{kaasbjerg2012} in \texttt{parameterKaasbjerg.hpp}) and the coupling comparison
of Fig.~\ref{fig:s3} are part of the dataset accompanying the main text.

\section{Adsorbate calculations}\label{sec:s3}

The sensor model of Sec.~4.2 of the main text takes the charge transferred per
adsorbed molecule as the amplitude of the transduction and the binding energy through
the adsorption constant. We computed both quantities for NO$_2$, NH$_3$ and O$_2$ on the
monolayer, with the same code, functional, pseudopotentials and lattice constant as the
electron-phonon calculations of Note~S2. Each molecule was placed 2.9 to 3.0~\AA{} above
the sulphur plane of a $4\times4$ supercell (48 atoms, one molecule per
$7.2\times10^{13}$~cm$^{-2}$, molecule images 12.7~\AA{} apart), on three sites, above a
Mo atom, above an S atom and above the hollow site. All atoms were then relaxed to forces
below $2\times10^{-3}$~Ry/bohr, with the Grimme-D3 dispersion correction, the
two-dimensional Coulomb cutoff, a $3\times3\times1$ mesh, a 60~Ry cutoff and spin
polarisation with the doublet (NO$_2$) and triplet (O$_2$) magnetisations. NO$_2$ was started with its oxygen
atoms towards the surface. Two further starting orientations, nitrogen towards the
surface and the molecule tilted with one oxygen down, test whether the flat orientation
is the minimum. The adsorption energy is
$E_{\text{ads}}=E(\text{layer+molecule})-E(\text{layer})-E(\text{molecule})$ with all
three terms in the same cell and settings, so that it is negative for a bound molecule.
The Langmuir kinetics of Sec.~4.2 of the main text use the binding energy
$E_{\text{b}}=-E_{\text{ads}}$, which is positive for a bound molecule. Charge transfer is taken as the Bader charge~\cite{henkelman2006} of the molecule in the relaxed complex minus its valence
charge. The L\"owdin charges of the projected wavefunctions were also computed and are
listed for reference, but they assign 0.1 to 0.2~$e$ more to every molecule than the
Bader partition because of the spilling of the projection and are not used. A
$5\times5$ cell (75 atoms, $2\times2\times1$ mesh) repeats the NH$_3$ case above Mo and the
NO$_2$ case above S.

\begin{table}[ht]
\caption{Adsorption energies and charge transfers of NO$_2$, NH$_3$ and O$_2$ on the
pristine monolayer (PBE-D3, $4\times4$ cell unless stated, and $\Delta q>0$ means electrons
transferred to the molecule). $h$ is the height of the lowest atom of the molecule
above the sulphur plane after relaxation. Published PBE values without dispersion
correction are NO$_2$ 0.100~$e$ and NH$_3$ 0.069~$e$ on the monolayer~\cite{yue2013},
0.097 and 0.009~$e$ on the bilayer~\cite{late2013}.}
\label{tab:s9}
\small
\begin{tabular}{@{}llllll@{}}
\toprule
Molecule & site & $E_{\text{ads}}$ (eV) & $h$ (\AA) & $\Delta q$ Bader ($e$) & $\Delta q$ L\"owdin ($e$)\\
\midrule
NO$_2$ & Mo & $-0.173$ & 3.03 & $+0.039$ & $-0.209$\\
NO$_2$ & S & $-0.175$ & 2.99 & $+0.048$ & $-0.200$\\
NO$_2$ & S, $5\times5$ & $-0.177$ & 2.98 & $+0.048$ & $-0.200$\\
NO$_2$ & S, started tilted & $-0.174$ & 2.94 & $+0.046$ & $-0.202$\\
NO$_2$ & S, N towards surface & $-0.130$ & 3.16 & $+0.031$ & $-0.214$\\
NO$_2$ & hollow & $-0.176$ & 3.01 & $+0.042$ & $-0.207$\\
NO$_2$ & hollow, N towards surface & $-0.150$ & 2.98 & $+0.022$ & $-0.224$\\
NH$_3$ & Mo & $-0.163$ & 2.91 & $-0.026$ & $-0.173$\\
NH$_3$ & Mo, $5\times5$ & $-0.168$ & 2.91 & $-0.027$ & $-0.174$\\
NH$_3$ & S & $-0.101$ & & $-0.012$ & $-0.169$\\
NH$_3$ & hollow & $-0.168$ & 2.91 & $-0.027$ & $-0.175$\\
O$_2$ & Mo & $-0.095$ & 3.21 & $+0.014$ & $-0.103$\\
O$_2$ & S & $-0.079$ & & $+0.006$ & $-0.107$\\
O$_2$ & hollow & $-0.093$ & 3.30 & $+0.011$ & $-0.105$\\
\bottomrule
\end{tabular}
\end{table}

All three molecules physisorb (Table~\ref{tab:s9}). They sit 2.9 to 3.3~\AA{} above
the sulphur plane, their adsorption energies differ between the sites by less than 5~meV
for NO$_2$ and by at most 70~meV for NH$_3$ and O$_2$, and the transferred charge is
small, 0.04 to 0.05~$e$ to NO$_2$, 0.01 to 0.03~$e$ from NH$_3$ and about 0.01~$e$ to
O$_2$. For NO$_2$, the flat
orientation with the oxygen atoms towards the surface is the minimum, since the tilted start
relaxes back to it, and the orientation with nitrogen towards the surface is a local
minimum 25 to 45~meV higher with a smaller transfer of 0.02 to 0.03~$e$. The $5\times5$
cell changes the NH$_3$ and NO$_2$ values by less than 2~meV and 0.001~$e$, so the
$4\times4$ results are converged with respect to the lateral separation of the
molecules. The NO$_2$ transfer is
half the published PBE value obtained without dispersion correction and with a
different site, and the NH$_3$ transfer lies at the low end of the published range.
Section~\ref{sec:s3-cmp} compares these values with the amplitudes of the sensor model.
The inputs and outputs of these calculations are part of the dataset.

\subsection{Comparison with the sensor-model amplitudes}\label{sec:s3-cmp}

Table~\ref{tab:s11} compares the fitted amplitudes of the sensor model with the computed
transfers and adsorption energies. Expressed per molecule with the site density of one
molecule per $4\times4$ cell ($7.2\times10^{13}$~cm$^{-2}$), the fitted amplitudes
correspond to 0.030~$e$ for NO$_2$ and 0.28~$e$ for NH$_3$. For NO$_2$, the fitted value
is two thirds of the computed transfer on the pristine plane, and the remaining
difference is the carrier limit of a device. The channel holds
$3\times10^{12}$~cm$^{-2}$ electrons, and once the acceptor level is filled to the point
where the Fermi level drops below it, further adsorption transfers no charge. The fit
obeys this bound ($\Delta N\theta\le n_0$), and the gate-voltage dependence of the
transfer computed by Late~\etal{}~\cite{late2013} shows the same limit. For NH$_3$, the
fitted value exceeds the
pristine-surface transfer ten-fold, which suggests a substantial contribution from
defect-related adsorption sites, where DFT gives much larger transfers, consistent with the OFF-state operating point of the
NH$_3$ measurements (Sec.~5.6 of the main text).

The binding energies lead to the same conclusion. Under the kinetic prefactor assumed
there, Eq.~(12) of the main text makes the fitted adsorption constants correspond to
effective binding energies $E_{\text{b}}$ of 0.47 and 0.60~eV for NO$_2$ on the devices of 
Kumar~\etal{}~\cite{kumar2025} and
Liu~\etal{}~\cite{liu2014}, respectively, and 0.47~eV for NH$_3$ on the device of
Liu~\etal{}~\cite{liu2014}. These lie between the physisorption binding energies of the pristine
basal plane (0.17 to 0.18~eV here, 0.28 and 0.25~eV for NO$_2$ and NH$_3$ in
Ref.~\cite{yue2013}) and the binding energies reported for defect-modified surfaces,
0.81~eV for NH$_3$ on nitrogen-incorporated MoS$_2$ carrying a sulphur
vacancy~\cite{kashyap2025} and 3.18~eV for NO$_2$ at a sulphur
vacancy~\cite{kushwaha2024}. The enhancement at a bare sulphur vacancy is not universal
in the literature. Regis~\etal{}~\cite{regis2024} keep NO$_2$ and NH$_3$ physisorbed at 0.05
to 0.30~eV whether or not a vacancy is nearby, and find only NO chemisorbed in the
vacancy, so the sites responsible for the stronger binding inferred here are not
identified by the fit. $K$ should therefore be interpreted as an effective adsorption
constant representing the unresolved distribution of adsorption sites, namely the basal
plane, vacancies and grain boundaries, whose weights differ from sample to sample. Its device
dependence therefore reflects the defect density, the growth method and the ambient of
each device rather than a deficiency of the transduction model, and the same Arrhenius
factor governs the temperature dependence of the recovery times reported for MoS$_2$
sensors.

\begin{table}[ht]
\caption{Fitted charge-transfer amplitudes of the sensor model of Sec.~4.2 of the main
text compared with the Bader charge transfers per molecule computed for this work
(PBE-D3, $4\times4$ cell, range over three sites, Table~\ref{tab:s9}) and with published
values.}
\label{tab:s11}
\begin{tabular}{@{}lll@{}}
\toprule
Quantity & NO$_2$ (acceptor) & NH$_3$ (donor)\\
\midrule
$\Delta N$ at full coverage (fitted) & $2.2\times10^{12}$~cm$^{-2}$ & $2.0\times10^{13}$~cm$^{-2}$\\
$N_c^{\max}$ (fitted) & $3.0\times10^{12}$~cm$^{-2}$ & $2.0\times10^{11}$~cm$^{-2}$\\
Effective charge per molecule (one per $4\times4$ cell) & 0.030~$e$ & 0.28~$e$\\
Bader charge, this work (pristine plane) & 0.039 to 0.048~$e$ & 0.012 to 0.027~$e$\\
$E_{\text{ads}}$, this work (pristine plane) & $-0.17$ to $-0.18$~eV & $-0.10$ to $-0.17$~eV\\
Effective binding energy $E_{\text{b}}=-E_{\text{ads}}$ of the fitted $K$ & 0.47 and 0.60~eV & 0.47~eV\\
Bader charge, monolayer, PBE~\cite{yue2013} & 0.100~$e$ & 0.069~$e$\\
Bader charge, bilayer, PBE~\cite{late2013} & 0.097~$e$ & 0.009~$e$\\
\bottomrule
\end{tabular}
\end{table}

\section{Extrinsic scattering forms, degenerate carrier statistics and reference comparison}\label{sec:s4}

\subsection{Extrinsic mechanisms}\label{sec:s4-extr}

The extrinsic mechanisms of the transport model (Table~1 of the main text) follow Ma and
Jena~\cite{majena2014} and Ridley~\cite{ridley1982}. Charged impurities at the interface scatter through a Thomas-Fermi- and
Rytova-Keldysh-screened 2D Coulomb potential,
\begin{equation}
V(q)=\frac{Z e^2}{2\varepsilon_0\varepsilon_{\text{avg}}}\,
\frac{e^{-q d}}{q_s+q+r_0 q^2},
\qquad q=2k\sin(\theta/2),
\label{eq:s-impurity}
\end{equation}
where $q_s$ is the free-carrier screening wavevector, $r_0$ the Rytova-Keldysh
length and $d$ the impurity setback. The polar substrate contributes remote
surface-optical (SO) phonon scattering, whose coupling scales with the dielectric
discontinuity $\tfrac{1}{2}[(\varepsilon_\infty^{\text{sub}}+\varepsilon_{\text{env}})^{-1}-(\varepsilon_0^{\text{sub}}+\varepsilon_{\text{env}})^{-1}]$
split between the substrate SO modes~\cite{majena2014}. Interface-roughness
scattering is included in the Prange-Nee form.
The adsorbate Coulomb centres of the sensor model are described by the same potential with
the areal density $N_c(\theta)$ of Eq.~(11) of the main text. The Rytova-Keldysh length
of the transport model is 40~\AA, within 10\,\% of the value $2\pi\alpha_{2D}=43.5$~\AA{} that
follows from the computed two-dimensional polarisability (Sec.~5.3 of the main text).

\subsection{Degenerate carrier statistics in the Poisson equation}\label{sec:s4-fd}

Because accumulation-mode two-dimensional channels reach degeneracy, the equilibrium
density in Poisson's equation (Eq.~(7) of the main text) follows Fermi-Dirac rather than
Boltzmann statistics. The local density and its derivative are
\begin{equation}
n(\phi)=\frac{1}{\beta}\ln\!\bigl(1+\beta\,e^{\,\varphi}\bigr),
\qquad
\frac{\partial n}{\partial\varphi}=\frac{e^{\,\varphi}}{1+\beta\,e^{\,\varphi}},
\qquad \varphi=\frac{q\phi-E_c+E_F}{k_BT},
\label{eq:s-fd}
\end{equation}
with the degeneracy parameter $\beta$ set by the two-dimensional band-edge density, and
$\beta\to0$ recovers the Boltzmann limit exactly, so non-degenerate materials are
unaffected. The derivative is used in the local linearisation of the nonlinear
Poisson update, and in the direct device simulation of Sec.~4.1 of the main text these statistics are
applied on the sheet nodes only.

\subsection{Comparison with the reference parametrization}\label{sec:s4-ref}

The LDA parametrization of Kaasbjerg~\etal{}~\cite{kaasbjerg2012} gives
$\Xi_{\text{LA}}=2.8$ and $\Xi_{\text{TA}}=1.6$~eV, first-order couplings of 5.9, 3.9
and 1.9~eV, homopolar and K-K$'$ couplings of $4.1\times10^{8}$ and
$2.6\times10^{8}$~eV/cm, and a Fr\"ohlich constant of 97~meV. Against these, the
computed LA coupling is twice as large and the TA coupling smaller, the first-order TA
and TO channels are two to three times larger, the zero-order optical couplings are 0.65
to 0.8 times as large, and the polar coupling is larger and wavevector dependent.
The first-order K-K$'$ couplings of the transport model are 12.0~eV (TA, 22.5~meV), 2.0~eV (LA,
28.5~meV), 4.8~eV (TO, 45.7~meV) and 3.5~eV ($A_1'$-like, 47.4~meV), and the intravalley
TO($\Gamma$) coupling 3.6~eV (Table~\ref{tab:s5b}). Table~5 of the main text shows that
the LA deformation potential alone accounts for the difference between the mobility obtained with the reference parametrization and with the computed couplings.

\section{Comparison with few-layer devices}\label{sec:s5}
\begin{figure}[ht]\centering
\includegraphics[width=\textwidth]{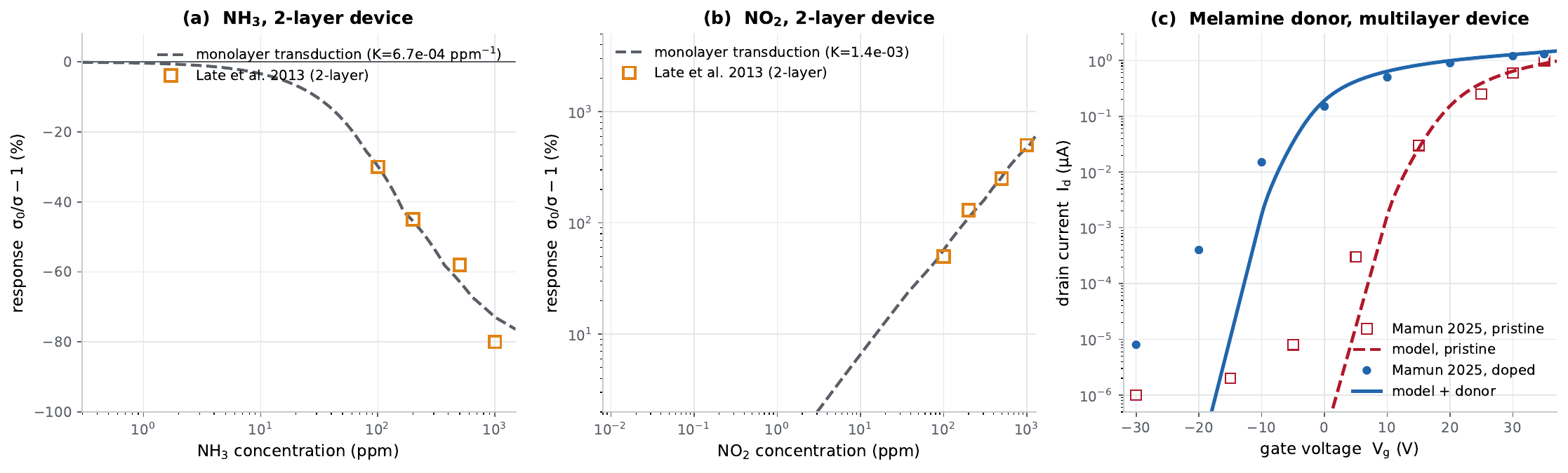}
\caption{Qualitative comparison with few-layer devices, not validation. (a) NH$_3$ and
(b) NO$_2$ response of a 2-layer device~\cite{late2013} against the monolayer
transduction with a fitted adsorption constant. (c) Transfer characteristics of a
multilayer FET before and after melamine doping~\cite{mamun2025}, reproduced by the
reduced device model as a donor shift of $\Delta V_{\text{th}}\approx-21$~V.}
\label{fig:s7}
\end{figure}

The framework of the main text is a monolayer model, and its quantitative claims are
monolayer claims. Few-layer and multilayer devices are nevertheless the majority of the
published sensing data, so Fig.~\ref{fig:s7} compares the monolayer transduction with a 2-layer
NH$_3$ and NO$_2$ device~\cite{late2013} and a multilayer melamine-doped
FET~\cite{mamun2025}, as a qualitative comparison rather than validation. The direction
of the response is expected to be the same in a few-layer film, because adsorbates act
only on the top surface,
and charge transfer and adsorbate Coulomb scattering act on the layers within the
screening length of MoS$_2$, which is a few layers~\cite{das2013,castellanos2013}. For a
bilayer, the entire film lies within that length, so the monolayer $\sigma(\theta)$ gives a
qualitative reference, with $K$ and $\Delta N$ absorbing the layer dependence of adsorption and
screening. The first-principles NO$_2$ transfer differs by 3\,\% between the monolayer and the
bilayer (0.100 and 0.097~$e$~\cite{yue2013,late2013}),
whereas the two NH$_3$ values differ by an order of magnitude (Table~\ref{tab:s11}),
which is why the NH$_3$ comparison is only qualitative. In thick films, the back gate
acts on the bottom layers and the adsorbate on the top layers, the interlayer resistance
network decouples the two, and the gate-tunable selectivity reported for few-layer
devices~\cite{late2013} is an interlayer effect outside the model. An extension to
few-layer films requires a layer-resolved model with interlayer coupling and screening
and lies outside the scope of this work.

\end{document}